\documentclass[amssymb,prb,twocolumn,floats,amsmath,showpacs,superscriptaddress]{revtex4}
\usepackage{bm}
\usepackage{graphicx}

\begin{document}
\preprint{Warsaw, August 11, 2005}

\title{Effect of the s,p-d exchange interaction on the excitons in (Zn,Co)O epilayers}

\author{W.~Pacuski}
\email{Wojciech.Pacuski@fuw.edu.pl}

\affiliation{Institute of Experimental Physics, Warsaw University,
Ho\.za 69, PL-00-681 Warszawa, Poland}

\affiliation{CNRS-CEA-UJF Joint Group "Nanophysique et
semiconducteurs", Laboratoire de Spectrom\'etrie Physique,\\BP 87,
F-38402 Saint Martin d'H\`eres Cedex, France}

\author{D.~Ferrand}
\affiliation{CNRS-CEA-UJF Joint Group "Nanophysique et
semiconducteurs", Laboratoire de Spectrom\'etrie Physique,\\BP 87,
F-38402 Saint Martin d'H\`eres Cedex, France}

\author{J.~Cibert}
\affiliation{CNRS-CEA-UJF Joint Group "Nanophysique et
semiconducteurs", Laboratoire de Spectrom\'etrie Physique,\\BP 87,
F-38402 Saint Martin d'H\`eres Cedex, France}
\affiliation{Laboratoire Louis N\'eel, CNRS, BP 166, F-38042
Grenoble Cedex 9, France}

\author{C.~Deparis}
\affiliation{Centre de Recherches sur l'H\'et\'ero\'epitaxie  et ses
Applications, CNRS,\\rue Bernard Gr\'egory, Parc Sophia Antipolis,
F-06560 Valbonne, France}

\author{J.~A.~Gaj}

\affiliation{Institute of Experimental Physics, Warsaw University,
Ho\.za 69, PL-00-681 Warszawa, Poland}

\author{P.~Kossacki}
\affiliation{Institute of Experimental Physics, Warsaw University,
Ho\.za 69, PL-00-681 Warszawa, Poland}
\affiliation{CNRS-CEA-UJF
Joint Group "Nanophysique et semiconducteurs", Laboratoire de
Spectrom\'etrie Physique,\\BP 87, F-38402 Saint Martin d'H\`eres
Cedex, France}

\author {C.~Morhain}
\affiliation{Centre de Recherches sur l'H\'et\'ero\'epitaxie  et ses
Applications, CNRS,\\rue Bernard Gr\'egory, Parc Sophia Antipolis,
F-06560 Valbonne, France}

\date{\today}

\begin{abstract}
We present a spectroscopic study of Zn$_{1-x}$Co$_{x}$O layers grown
by molecular beam epitaxy on sapphire substrates.
Zn$_{1-x}$Co$_{x}$O is commonly considered as a promising candidate
for being a Diluted Magnetic Semiconductor ferromagnetic at room
temperature. We performed magneto-optical spectroscopy in the
Faraday configuration, by applying a magnetic field up to 11~T, at
temperatures down to 1.5~K. For very dilute samples ($x<0.5\%$), the
giant Zeeman splitting of the $A$ and $B$ excitons is observed at
low temperature. It is proportional to the magnetization of isolated
Co ions, as calculated using the anisotropy and g-factor deduced
from the spectroscopy of the ${d\texttt{-}d}$ transitions. This
demonstrates the existence of spin-carrier coupling. Electron-hole
exchange within the exciton has a strong effect on the giant Zeeman
splitting observed on the excitons. From the effective spin-exciton
coupling, ${\langle N_0 (\alpha-\beta) \rangle_X=0.4}$~eV, we
estimate the difference of the exchange integrals for free carriers,
${N_0 |\alpha - \beta|\simeq0.8}$~eV. The magnetic circular
dichroism observed near the energy gap was found to be proportional
to the paramagnetic magnetization of anisotropic Co ions even for
higher Co contents.
\end{abstract}

\draft



\maketitle


\section {Introduction}
Theoretical predictions,\cite{Sato00,Diet01,Blin02,Spal04} and first
reports of room temperature ferromagnetism in
Zn$_{1-x}$Co$_{x}$O,\cite{Ueda01,Lim03,Rode03,Rama04,Tuan04,Kitt05a,Venk04}
have strongly stimulated researches on this wide bandgap Diluted
Magnetic Semiconductor (DMS). However, the origin of the observed
ferromagnetic behavior remains controversial, and the existence of a
coupling between the magnetic and electronic properties has still to
be demonstrated, and its magnitude evaluated.\cite{Venk04}

Most of reported studies rely on magnetometry. Direct magnetization
measurements on thin DMS layers are particularly difficult, because
the magnetic signal from the thin layer is usually small compared to
the diamagnetic contribution from the much thicker substrate.
Magneto-optical spectroscopy on the excitonic transition and
Magnetic Circular Dichroism (MCD) in the vicinity of the bandgap can
be used very efficiently to measure the temperature and magnetic
field dependence of the magnetization, using the giant Zeeman effect
of substitutional magnetic ions coupled to the band electrons, and
to discriminate the contribution from magnetic ions incorporated in
spurious phases. In addition, the observation of the intra-ionic
${d\texttt{-}d}$ transitions gives us an information on the
electronic structure of the incorporated magnetic impurity, and
allows us to measure accurately the parameters describing its ground
level and governing the magnetization of the isolated impurities.
Finally, combining the two sets of data, on can deduce the magnitude
of the ${s,p\texttt{-}d}$ exchange interactions between localized
spins and free carriers.

This paper describes the results of such a study conducted on a set
of Zn$_{1-x}$Co$_{x}$O layers with various Co concentrations,
including very dilute samples which feature well resolved excitonic
lines, so that we can measure the exchange Zeeman splitting. It
fills the gap existing between recent magneto-optical studies of
Zn$_{1-x}$Co$_{x}$O layers with a high Co content, where large
linewidths prevent any direct observation of the exchange Zeeman
splitting,\cite{Ando01,Ando01b,Ando02,Ando04,Schw03,Kitt05b,Tuan04}
and early studies of very dilute Zn$_{1-x}$Co$_{x}$O bulk samples
without applied field.\cite{Koid77,Papp61,Anderson67,Shul87}


\section {Samples and experiment}

About 1$\mu$m-thick layers were grown on sapphire substrates by
plasma-assisted molecular beam epitaxy. Two dimensional growth was
achieved for a growth temperature of 560$^\circ$C (i.e., 50$^\circ$C
higher than the optimal growth temperature used for ZnO), resulting
in streaky RHEED patterns. The Co content of several layers was
measured by energy dispersive x-ray analysis (EDX). For low Co
contents (${x < 1\%}$), the full widths at half maximum of the x-ray
rocking curves are in the range of ${\omega\sim 0.15^\circ}$ along
(002), ($\overline 105$), and (105). The low, identical values of
$\omega$ measured both for ($\overline105$) and (105), indicate a
large column diameter, close to 1$\mu$m (up to ${x=15\%}$). While
the column diameters remain large, $\omega$ values are found to
increase slightly and gradually with the Co concentration. For ${x=
15\%}$, measured $\omega$ values are $0.32^{\circ}$, $0.22^{\circ}$,
and $0.66^{\circ}$ along (002), ($\overline105$), and (101)
respectively. For all compositions, the c-axis of the wurtzite
structure is perpendicular to the film plane. No other orientation
or column rotation were detected. The conductivity of the films is
\emph{n}-type, with residual carrier concentrations $n_e$~$<
1.10^{18}$~cm$^{-3}$, below the Mott transition. The thickness of
the layer was checked on the electron microscope image of a cleavage
plane.

Reflectivity, transmission and photoluminescence (PL) measurements
were performed in the Faraday configuration, with the optical axis
and a magnetic field (up to 11~T) both parallel to the c-axis, at
temperatures down to 1.5~K. A high-pressure Xe lamp was used for
transmission and reflectivity; PL was excited with a He-Cd laser.


\section {Spectroscopy of cobalt ions}
\label{sec:Co}

We measured the characteristic absorption lines and bands in every
studied Zn$_{1-x}$Co$_{x}$O sample from the most diluted one
($x=0.1\%$) up to the most concentrated one (${x=35\%}$). Examples
are given in Fig. \ref{fig:Cocalibration}(a) for the spectral range
of interest in the following. These lines and bands have been
already identified as intra-ionic ${d\texttt{-}d}$ transitions of
cobalt in bulk samples by P.~Koidl.\cite{Koid77} For substitutional
Co ions, the different eigenstates associated to the 3d$^7$
configuration are labeled using the notation of
Mcfarlane.\cite{Macf70}

We will particularly study the transitions between the $^4A_2$
ground state and the $^2E$ excited state arising from the cubic part
of the crystal field. Both quadruplets are further split into two
components (two Kramers doublets), $2\overline A$ and $\overline E$,
by the spin-orbit coupling and the trigonal component of the crystal
field. For the $^4A_2$ ground states, we label
$|$$\pm$$\frac12\rangle$ the two spin sublevels of $\overline E$ and
$|$$\pm$$\frac32\rangle$ the two spin sublevels of $2\overline A$.
The ${\overline E-2\overline A}$ splitting of the $^4A_2$ ground
states is to small to be resolved in the spectra of
Fig.~\ref{fig:Cocalibration}(a), while that of the $^2E$ excited
state is clearly visible.

Other lines which can be seen at higher energy in
Fig.~\ref{fig:Cocalibration}(a) have been attributed by Koidl
\cite{Koid77} to transitions to other excited levels. They are
broader, and hence more difficult to study quantitatively in thin
layers since transmission spectra are plagued by interferences
between the surface of the layer and the interface with the
substrate.

In this section we use the $^4A_2\leftrightarrow\,^2E$ transition to
(A) estimate the Co content in samples which have not been
characterized by EDX (particularly at low Co content), (B) confirm
the parameters describing the ground state and the excited state
(anisotropy and g-factors), and the oscillator strengths of
transitions and (C) from a complete description of the intensities
of the absorption lines, confirm the population distribution in the
ground level and deduce the magnetization of the system of localized
spins.

\subsection {Calibration of Co concentration}

\begin{figure}
\includegraphics*[width=85mm]{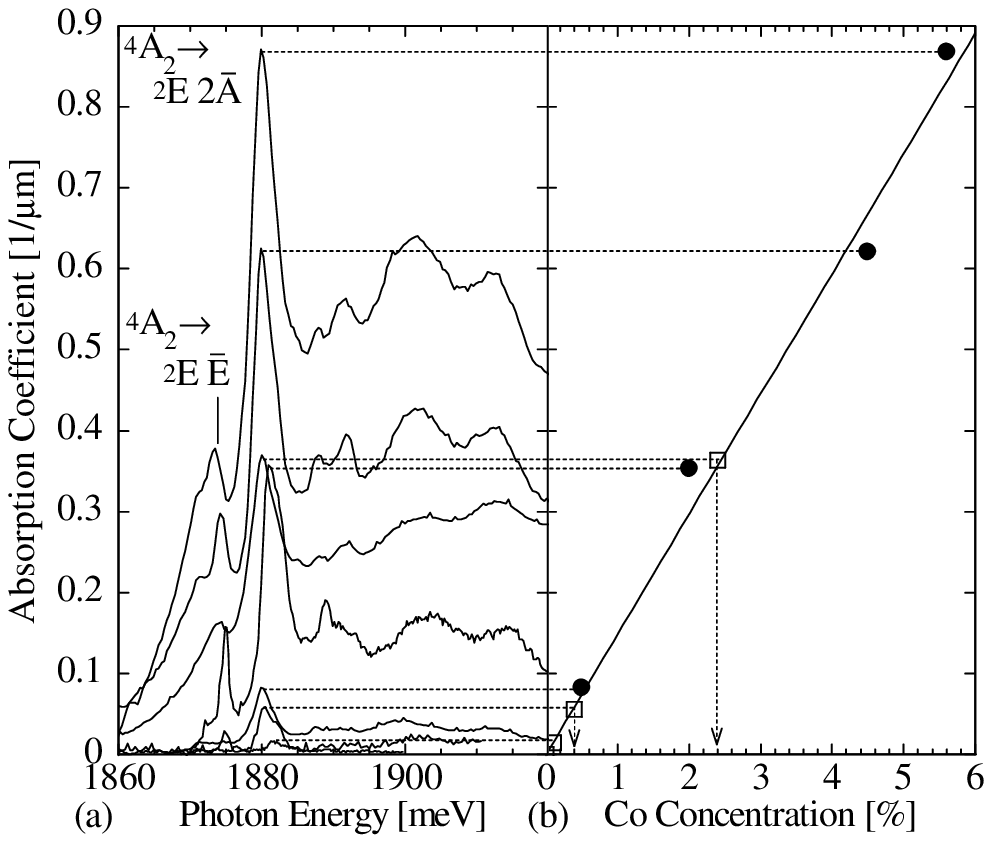}
\caption[]{(a) Absorption spectra of Co$^{2+}$ in
Zn$_{1-x}$Co$_{x}$O samples with various cobalt concentration $x$,
at ${T=1.6}$~K. Two zero-phonon intra-ionic transitions are
identified: $^4A_2 \rightarrow\,^2E \overline E$ (at 1875~meV) and
$^4A_2 \rightarrow\,^2E 2\overline A$ (at 1880~meV). The~absorption
coefficient of the $^4A_2 \rightarrow\,^2E 2\overline A$ transition
is shown by a dotted horizontal line pointing to the right figure.
 (b)~The~absorption coefficient at 1880~meV, as a function of the~Co
concentration determined by EDX (full circles). The observed linear
dependence allows us to determine also the Co concentration in very
diluted samples (open squares and vertical arrows), where the EDX
technique is not sensitive enough.} \label{fig:Cocalibration}
\end{figure}

In a set of samples, where the cobalt concentration was determined
by EDX, the absorption coefficient of the $^4A_2 \rightarrow\,^2E
2\overline A$ line [at 1880meV, Fig.~\ref{fig:Cocalibration}(a)]
increases linearly with the Co concentration up to 6\%. [full
circles in Fig.~\ref{fig:Cocalibration}(b)]. The calibration curve
obtained with these samples was used to determine the cobalt
concentration in other samples [open squares in
Fig.~\ref{fig:Cocalibration}(b)].

\begin{figure*}[t] 
\includegraphics*[width=180mm]{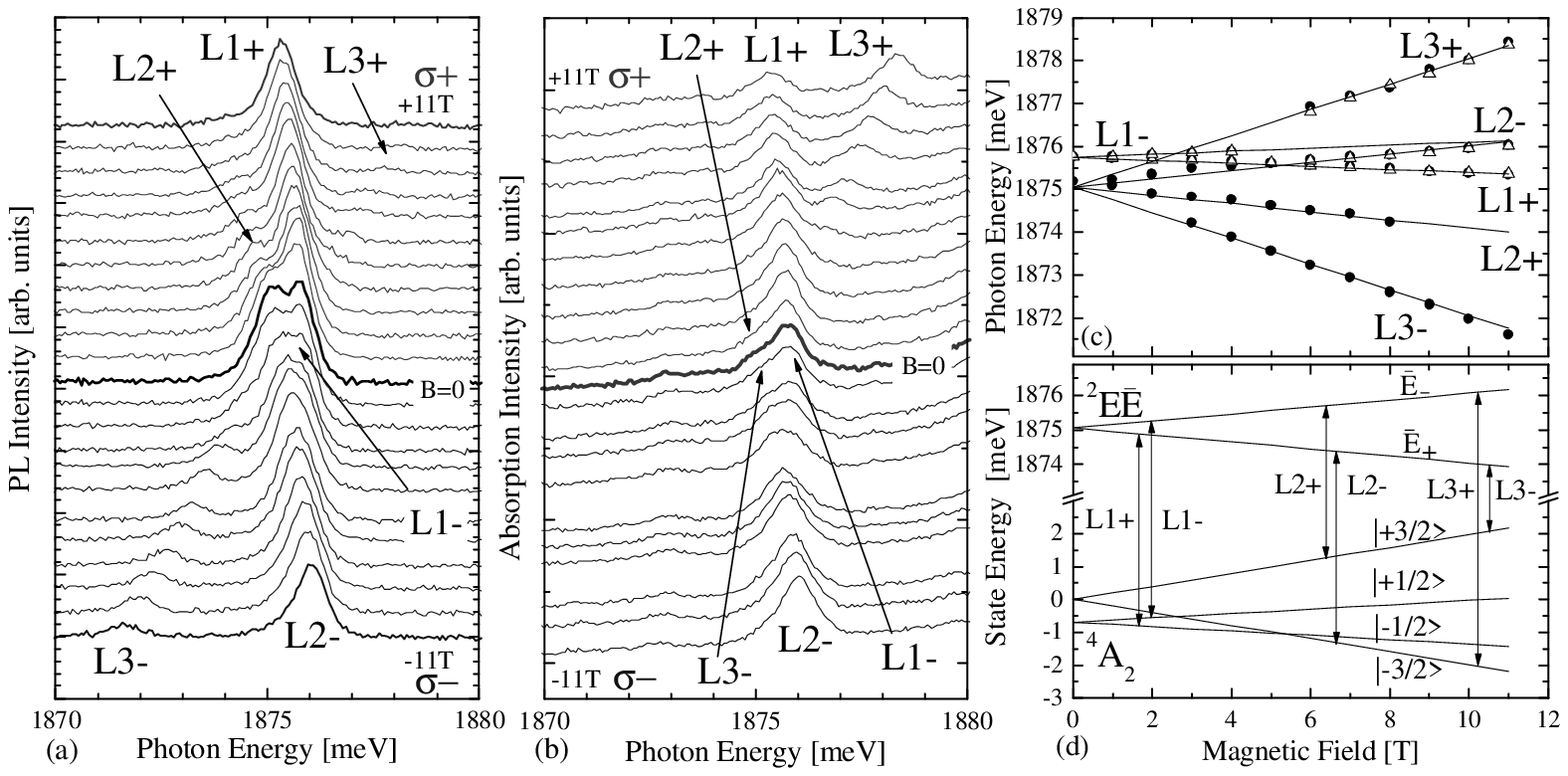}
\caption[]{Magneto-optical spectroscopy of the
$^4A_2\leftrightarrow\,^2E \overline E$ transition in a
Zn$_{0.98}$Co$_{0.02}$O at $T=7$~K, in $\sigma^{+}$ (top) and
$\sigma^{-}$ (bottom) circular polarizations. The magnetic field up
to 11~T is parallel to the c-axis and to the propagation of light
(Faraday configuration). (a) PL spectra. (b) Absorption spectra. (c)
Position in energy of the observed PL (full circles) and absorption
(triangles) lines, and (solid lines) values calculated using the
parameters of Table \ref{tab:PL}. (d) Energy level diagram, as a
function of the applied field, and transitions. The spin levels of
the ground level quadruplet are marked.} \label{fig:intra-ionic}
\end{figure*} 

The linear behavior observed up to a Co concentration of about 6\%
indicates that the absorption coefficient is sensitive to the total
concentration of Co$^{2+}$ ions substituting Zn in the wurtzite
lattice. In particular, it does not discriminate the different
nearest neighbor configurations: isolated ions, antiferromagnetic or
ferromagnetic cobalt pairs, triplets, or other complexes. In the
cation sublattice of the wurtzite structure, every cation has 12
nearest neighbors: 6 neighbors in the same layer, and 3 neighbors in
the next layer on each side, at almost the same distance. These 12
neighbors are usually considered as equivalent (see, e.g., the study
of Zn$_{1-x}$Mn$_{x}$O\cite{chik05}). Assuming a random distribution
of Co ions, the probability to find an isolated cobalt ion (without
any cobalt ion in a nearest neighbor position) is given by
${(1-x)^{12}}$. For a total concentration $x=5.6\%$, half of the Co
ions have at least one Co ion as a nearest neighbor. That means that
the concentration of isolated Co ions, ${x_s= x(1-x)^{12}}$,
significantly deviates from the straight line drawn in
Fig.~\ref{fig:Cocalibration}(b)].

The magnetic properties of a set of isolated Co ions and of nearest
neighbors complexes are expected to be very different at low
temperature. Antiferromagnetic nearest-neighbor pairs are blocked
antiparallel; in many DMSs, this gives rise to a series of steps and
plateaux in the high field part of the magnetization cycles. At
smaller fields, the magnetization is well described if one replaces
the total number of spins, $x$, by an effective number of spins,
$x_{eff}$. The value of $x_{eff}$ has been determined in a large
number of DMSs with the zinc-blende structure. It does not depend on
the chemical nature of the DMS, which suggests that it corresponds
to a statistical distribution in an ensemble of randomly occupied
sites. For small values of $x$, the above value of $x_s$ coincides
with the experimental value\cite{Gaj94} of $x_{eff}$ in
Cd$_{1-x}$Mn$_{x}$Te. We shall use this approximation in the
following.

\subsection {Magnetospectroscopy}
We combine PL and absorption data involving the
$^4A_2\leftrightarrow\,^2E\overline E$ intra-ionic transitions in
order to check the parameters governing the evolution under magnetic
field of the $^4A_2$ ground state. Fig.~\ref{fig:intra-ionic}(a)
shows the spectra around the zero phonon PL lines near 1875 meV. Two
PL lines with equal intensities are observed in zero magnetic field.
They have been identified\cite{Koid77} as transitions from the
excited state $^2E \overline E$ to the two components of the ground
state ($|$$\pm$$\frac12\rangle$ and $|$$\pm$$\frac32\rangle$). In
the presence of a magnetic field parallel to the c-axis, we observe
six PL lines [see Fig. \ref{fig:intra-ionic}(a)], which were
identified as transitions between the different spin sublevels of
the ground and excited states split by the Zeeman effect. In
agreement with the selection rules for dipolar transitions (see
Table~\ref{tab:rules}),\cite{Macf70} for light propagating along the
c-axis, two transitions are forbidden, three lines are observed in
$\sigma^+$ polarization and three in $\sigma^-$ polarization.

\begin{table}
\begin{ruledtabular}
\begin{tabular}{ccccc}
\vspace{0.1cm}
&$|\texttt{+}\frac12\rangle$ & $|\texttt{-}\frac12\rangle$ & $|\texttt{+}\frac32\rangle$ & $|\texttt{-}\frac32\rangle$ \\
\vspace{0.1cm}
&  $^4A_2 \overline E_+$&  $^4A_2 \overline E_-$&  $^4A_2\overline A  $&  $^4A_2 \overline A  $\\

\hline
$^2E \overline E_+\, (|\texttt{+}\frac12\rangle)$&$\pi$     &$\sigma^+$&$\sigma^-$&$\sigma^-$\\

$^2E \overline E_-\, (|\texttt{-}\frac12\rangle)$&$\sigma^-$&$\pi$&     $\sigma^+$&$\sigma^+$\\

\end{tabular}
\end{ruledtabular}
\caption{\label{tab:rules}Optical selection rules for dipole
transitions in trigonal symmetry (see Macfarlane\cite{Macf70}).
These selection rules directly reflect the conservation of total
momentum, modulo 3. The $\pi$ polarization corresponds to the active
field of the light parallel to the c-axis of the wurtzite crystal
and cannot be observed when light propagates along the c-axis; the
$\sigma$ polarization corresponds to the active field perpendicular
to the c-axis, and is observed with both helicities when light
propagates along c.}
\end{table}

The transitions observed in PL are also visible in absorption
[Fig.~\ref{fig:intra-ionic}(b)]. Fig.~\ref{fig:intra-ionic}(c) shows
the position of the PL and absorption lines as a function of the
intensity of the magnetic field up to 11T. The solid lines are
calculated using 4 fitting parameters: values of the Land\'e factors
${g_{||}=2.28}$ for the ground state and ${g_{||}'=-3.52}$ for the
excited state, a zero-field splitting of the ground state
${2D=0.69}$~meV, and an energy of the zero field transition between
$|$$\pm$$\frac32\rangle$ and $^2E\overline E$  equal to 1875~meV.
These parameters, obtained for a (strained) layer with 2\%~Co, are
in good agreement with the values obtained theoretically or from
other experimental techniques (see Table~\ref{tab:PL}). The
corresponding energy diagram and the scheme of transitions are shown
in Fig.~\ref{fig:intra-ionic}(d).

\begin{table} 
\begin{ruledtabular}
\vspace{0.1cm}
\begin{tabular}{ccccc}
\vspace{0.1cm} & Parameter &This work
&References\\
\hline
$^2E \overline E$ &$g_{||}'$&-3.52 &
-3.358\footnote[1]{Theory,
calculated by R. M. Macfarlane\cite{Macf70}}\\
$^4A_2$&$g_{||}$&2.28 & 2.2384\footnotemark[2]\\

$|$$\pm$$\frac12\rangle \leftrightarrow |$$\pm$$\frac32\rangle$&
$2D$&0.69 meV&0.68 meV\footnote[2]{EPR experiment done by N. Jedrecy
et al.\cite{Jedr04}, diluted
sample}\\

$|$$\pm$$\frac32\rangle \leftrightarrow\,^2E\overline E$& energy &
1875 meV & 1877 meV\footnote[3]{Optical absorption
measured by P. Koidl\cite{Koid77}, diluted sample}\\
\end{tabular}
\end{ruledtabular}
\caption{\label{tab:PL}Parameters describing the spin splitting of
the $^4A_2$ ground state quadruplet and the $^4A_2
\leftrightarrow\,^2E \overline E$ transition. The values were
derived from PL and absorption measurements of Zn$_{1-x}$Co$_{x}$O
with 2\%~Co.}
\end{table} 

Since we shall be interested in the magnetic properties of the Co
system, it is important to note that in zero magnetic field, the
$|$$\pm$$\frac12\rangle$ states are at lower energy than the
$|$$\pm$$\frac32\rangle$ states. This ordering induces a spin
anisotropy with an easy axis perpendicular to the c-axis. Due to the
three times larger Zeeman splitting\cite{Macf70} of the
$|$$\pm$$\frac32\rangle$ states, if the field is along the c-axis,
we observe a crossing of $|\texttt{-}\frac32\rangle$ with
$|\texttt{-}\frac12\rangle$ at 5.2~T [See
Fig.~\ref{fig:intra-ionic}(d)].

\begin{figure} 
\includegraphics*[width=85mm]{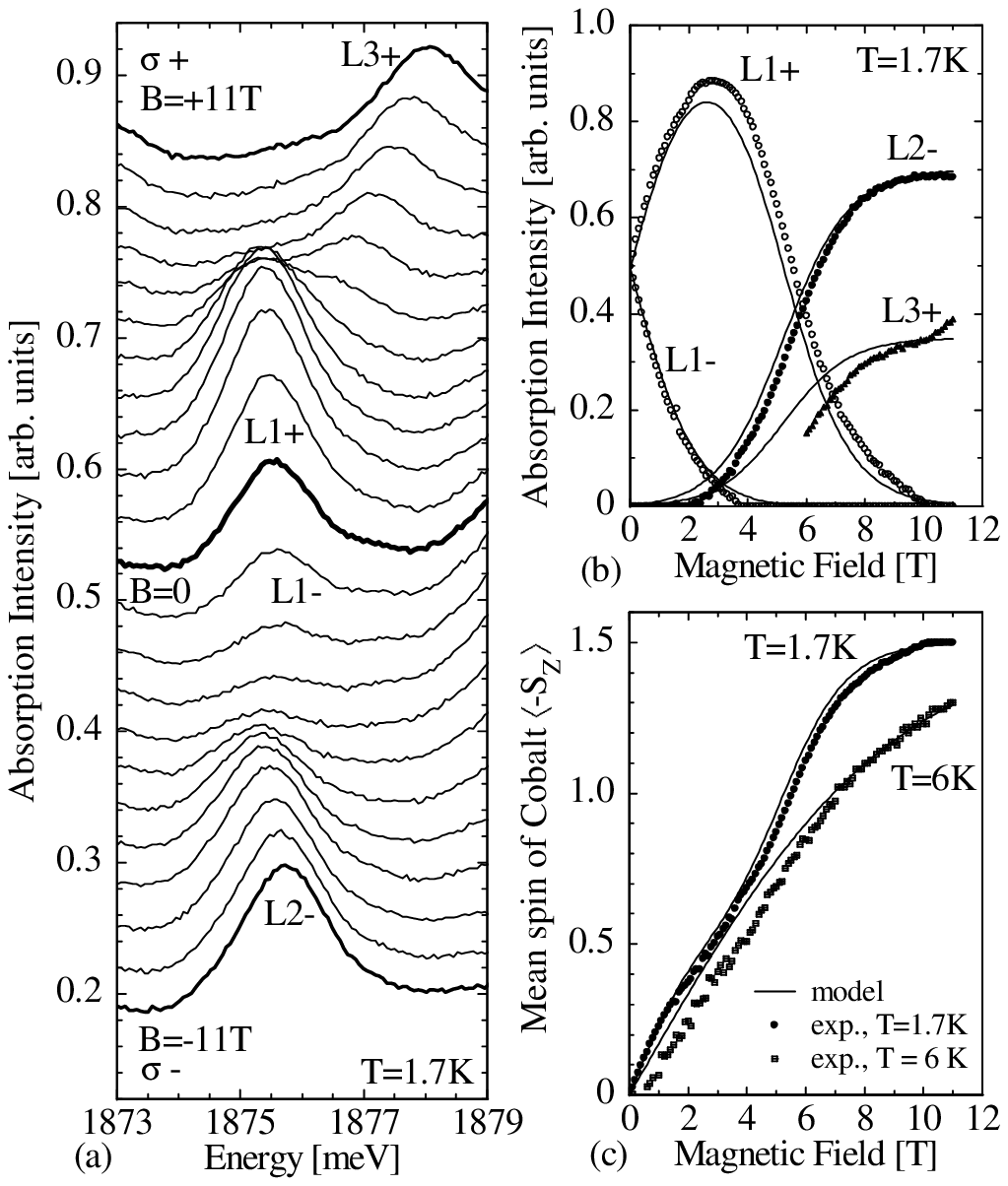}
\caption[]{(a) Absorption spectra of Zn$_{0.955}$Co$_{0.045}$O
measured in the Faraday configuration, at 1.7~K, in a magnetic field
up to 11~T, in $\sigma^{+}$ and $\sigma^{-}$ circular polarizations.
The four transitions observed are labeled according to Fig.
\ref{fig:intra-ionic}(d). (b) Absorption intensity measured at
$T=1.7$~K (symbols) and calculated using Maxwell-Boltzmann
statistics (lines). (c) Mean spin of Co calculated from the
absorption intensities (symbols) or from Maxwell-Boltzmann
statistics and parameters of Table \ref{tab:PL} (lines).}
\label{fig:abs-mag}
\end{figure} 

\subsection {Populations of the spin sublevels and magnetization}
The magnetization of the Co system is determined by the population
of the different spin sublevels of the ground state. We now check
these populations from the absorption intensities and calculate the
expected magnetization.

For a better accuracy, we used magneto-absorption spectra from a
sample with a higher Co content, 4.5\% [Fig.~\ref{fig:abs-mag}(a)].
As each absorption intensity is proportional to the population of
the initial state, we observe only four $^4A_2
\rightarrow\,^2E\overline E$ transitions, which are labeled as in
Fig.~\ref{fig:intra-ionic}.

Using the parameters determined from the previous PL and absorption
data (Table \ref{tab:PL}), and assuming gaussian line shapes with a
linewidth equal to 1.44~meV, we were able to fit the whole set of
spectra with the intensity of each line as adjustable parameters.
The resulting intensities are shown by symbols in
Fig.~\ref{fig:abs-mag}(b). At low temperature (${T=1.7}$~K), two
transitions (L2$+$ and L3$-$) keep zero intensity for every magnetic
field.

The integrated intensity of each absorption line is proportional to
the occupancy of the initial state and to the oscillator strength of
the transition. Relative oscillator strengths are easily determined
from extreme cases where one level is populated. At $T=1.7$~K and
${B=11}$~T only transitions L2$-$ and L3$+$ are visible
[Fig.~\ref{fig:abs-mag} (a) and (b)] because only  the
$|\texttt{-}\frac32\rangle$ state is occupied. Similarly, in zero
magnetic field only the $|\texttt{-}\frac12\rangle$ and
$|\texttt{+}\frac12\rangle$ states are occupied so that only L1$+$
and L1$-$ are observed. Measuring the relative intensities in zero
magnetic field and with ${B=11}$~T leads to the conclusion that the
relative oscillator strengths of L1$+$, L2$-$, L3$+$ are 1 : 0.7 :
0.35 [see Fig.~\ref{fig:abs-mag}(b)]. We expect that the pairs of
lines (L1$+$ and L1$-$), (L2$+$ and L2$-$), and (L3$+$ and L3$-$),
have equal oscillator strengths because of time reversal symmetry in
zero field. Independently, the relative oscillator strength of
L1$+$, L2$-$ and L3$-$ can be estimated from the intensity of PL
lines in $B=11$~T, when only the lower component ($\overline E_+$)
of $^2E\overline E$ excited state is occupied during the lifetime
[Fig.~\ref{fig:intra-ionic}(a)]. The oscillator strengths are
summarized in Table~\ref{tab:OS}.

\begin{table} 
\begin{ruledtabular}
\vspace{0.1cm}
\begin{tabular}{lcccccc}
\vspace{0.1cm}
 Transition& L$1+$ & L$1-$ & L2$+$ & L2$-$ & L3$+$ & L3$-$  \\
\vspace{0.1cm} Excited state&$\overline E_+$
 & $\overline E_-$ & $\overline E_-$
 & $\overline E_+$& $\overline E_-$
 & $\overline E_+$\\
 \vspace{0.1cm} Ground state&$|\texttt{-}\frac12\rangle$
 & $|\texttt{+}\frac12\rangle$ & $|\texttt{+}\frac32\rangle$
 & $|\texttt{-}\frac32\rangle$& $|\texttt{+}\frac12\rangle$
 & $|\texttt{-}\frac12\rangle$\\
\vspace{0.1cm} Relative OS (from abs.) & 1 & 1 & 0.7\footnote[1]{
from time reversal symmetry} & 0.7 & 0.35\footnote[2]{not very
accurate, see Fig. \ref{fig:abs-mag}(b)} &   \\
 \vspace{0.1cm} Relative OS (from PL) & 1 &  &  & 0.8 &  &  0.15\footnotemark[2]\\
\end{tabular}
\end{ruledtabular}
\caption{\label{tab:OS}Ratio of the oscillator strengths (OS) of the
spin split components of the $^4A_2 \leftrightarrow\,^2E 2\overline
A$ transition in $\sigma$ polarization. OS equal to 1 corresponds to
a peak intensity of 0.14/$\mu$m with a linewidth of 1.44~meV
measured on a sample with 4.5\%~Co.}
\end{table}

The occupancy of the different spin sublevels can be calculated
using the parameters of Table~\ref{tab:PL} and assuming a
Maxwell-Boltzmann distribution. The solid lines in
Fig.~\ref{fig:abs-mag}(b) show the corresponding intensity
calculated for each line using the relative oscillator strengths of
Table \ref{tab:OS}. Experimental data for L1$+$, L1$-$, and L2$-$
are in good agreement with the calculation. A significant
discrepancy appears for L3$+$. For this line, the determination of
the integrated intensity is made difficult by the overlap with
L1$+$. Moreover a magnetic field dependence of the oscillator
strength cannot be ruled out experimentally.

The experimental absorption intensities divided by the relative
oscillator strengths give us a direct information about the expected
cobalt mean spin, which can be calculated using
\begin{equation}
\langle \texttt-S_z\rangle_{spectro} = \frac{\frac12 (I_{1+} -
I_{1-}) +\frac32(I_{2-} - I_{2+})/r_{2,1}}{\frac12 (I_{1+} + I_{1-})
+\frac32(I_{2-} + I_{2+})/r_{2,1}},
 \label{eq:meanSz-exp}
\end{equation}
where $I_N$ denotes the experimental intensity of absorption line
L$_N$, and $r_{2,1}$ denotes the oscillator strength ratio of lines
L2$\pm$ and L1$\pm$ (it is equal to 0.7, see Table~\ref{tab:OS}). We
do not use L3$\pm$ since its experimental intensity is doubtful. The
expected value is shown by symbols in Fig.~\ref{fig:abs-mag}(c), as
a function of the magnetic field along the c-axis, for two
temperatures: 1.7~K and 6~K. It clearly deviates from an isotropic
Brillouin function.

Here again, assuming a Maxwell-Boltzmann distribution and using the
parameters of Table~\ref{tab:PL}, we can calculate the mean spin of
an isolated cobalt, as a function of temperature $T$ and magnetic
field $B_z$ along the c-axis,
\begin{subequations}
\label{eq:meanSz}
\begin{equation}
\langle \texttt-S_z\rangle = \frac{\frac12
\sinh(\frac12\delta)+\frac32
e^{-\frac{2D}{k_{B}T}}\sinh(\frac32\delta)}{\cosh(\frac12\delta)
+e^{-\frac{2D}{k_{B}T}}\cosh(\frac32\delta)},
\end{equation}
\begin{equation}
{\delta} = \frac{g_{||}\mu_B B_z}{k_{B}T},
\end{equation}
\end{subequations}
where $k_{B}$ denotes the Boltzmann constant and $\mu_B$ the Bohr
magneton. Parameters $g_{||}$ and 2$D$ are given in
Table~\ref{tab:PL}. The result is shown as solid lines in
Fig.~\ref{fig:abs-mag}(c).

Particularly for very low temperatures, the field dependence of the
magnetic moment deviates from the Brillouin function B$_{3/2}$
characteristic for an isotropic spin. A~step is visible at 5~T, as
an effect of the crossing between $|\texttt{-}\frac32\rangle$ and
$|\texttt{-}\frac12\rangle$ [Fig.~\ref{fig:intra-ionic}(d)]. Also,
if we plot the inverse of the low-field susceptibility as a function
of temperature, we observe a clear deviation from a Curie
law.\cite{Ferr05} These two effects mimic the magnetization steps
and Curie-Weiss law observed to be due to spin-spin interactions in
the case of DMSs containing the isotropic Mn spin; here they are
simply due to the single-ion anisotropy of the Co spin.

The magnetization $\langle M_z\rangle$ of the cobalt system is
directly related to the previous mean spin value
[Eqs.~(\ref{eq:meanSz-exp}) or (\ref{eq:meanSz})]:
\begin{equation}
\label{eq:magnetization} \langle M_z\rangle = -g_{||}~\mu_B~
N_0~x_{eff}~\langle S_z\rangle
\end{equation}
In section~\ref{sec:X} we describe results of magnetooptical
measurements performed at photon energies close to the energy gap of
Zn$_{1-x}$Co$_{x}$O. In section~\ref{sec:Analysis}, we compare them
with the magnetization above.


\section {Excitonic SPECTROSCOPY}
\label{sec:X}

In this section, we first recall the main characteristics of the
spectroscopy of ZnO, a wide gap semiconductor with the wurtzite
structure and a small spin-orbit coupling. Then we describe the
giant Zeeman effect observed on sharp excitonic transitions in very
dilute Zn$_{1-x}$Co$_{x}$O layers. Finally we turn to our
observations of near-gap magnetic circular dichroism in layers with
a larger cobalt content.
\begin{figure}
\includegraphics*[width=85mm]{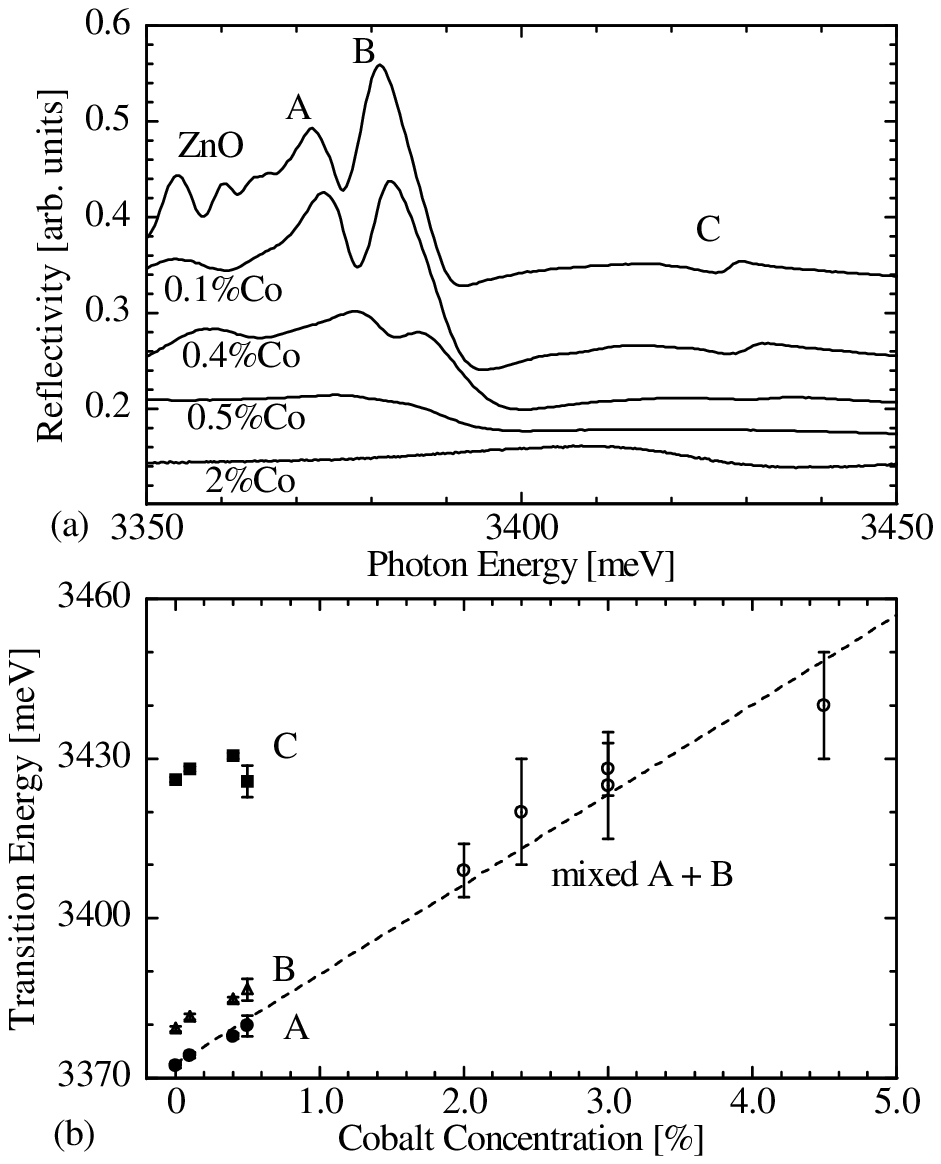}
\caption[]{(a)  Reflectivity spectra measured at $T=1.6$~K with
incidence angle of 45~degrees. Topmost spectrum is for ZnO, other
spectra for Zn$_{1-x}$Co$_{x}$O with increasing Co concentration
(0.1\%, 0.4\%, 0.5\%, 2\%). Labels $A$, $B$ and $C$ identify the
three excitons which are visible in ZnO and at low Co content;
Perot-Fabry oscillations are also observed at low photon energy. (b)
Spectral positions of the excitonic transition in ZnO and
Zn$_{1-x}$Co$_{x}$O. Symbols marked as "mixed $A$+$B$" means that in
the case of samples with a Co concentration higher than 2\% we
cannot resolve excitons $A$ and $B$, and we plot the position of a
broad structure. The dashed line shows the linear fit for the
position of exciton $A$ as a function of Co content.}
\label{fig:Reflectivity45}
\end{figure}

\subsection {Zero field results}
ZnO and Zn$_{1-x}$Co$_{x}$O naturally crystallize in the wurtzite
structure. In this structure, the trigonal component of the crystal
field and the spin-orbit coupling split the valence band in the
center of Brillouin zone into three valence band edges. These
splittings are small and, in epitaxial layers, they depend on the
residual strain. Actually, the position and the symmetry of the
valence edges are still a mater of controversy even in bulk ZnO.

Three excitons can be detected in ZnO. They are labeled $(A,B,C)$,
in the order of their position in the spectra, independently of
their symmetry. Exciton $A$, which appears at the lowest energy, is
associated to the valence band edge at the highest energy, and $C$
to the edge at the lowest
energy.\cite{Langer70,Reynolds99,Gil01a,Gil01b,Lamb02} Both the $A$
and $B$ excitons are observed by reflectivity at normal incidence
(${k\parallel c}$), hence in $\sigma$ polarization (electric field
of the light perpendicular to the c-axis of the crystal, ${E\perp
c}$), while exciton $C$ can be observed only in $\pi$ polarization
(${E\parallel c}$).

These three valence band edges result from the combined effect of
the trigonal crystal field (described by a parameter noted
$\Delta_1$) and the anisotropic spin-orbit coupling (two parameters,
${\Delta_2=\Delta_{SO}^{||}/3}$,
${\Delta_3=\Delta_{SO}^{\perp}/3}$). In ZnO, the spin-orbit coupling
is much smaller than the trigonal crystal field (more precisely,
$\Delta_3$ is much smaller than $\Delta_1 -\Delta_2$). Hence the
trigonal field splits the \emph{p}-like states which form the top of
the valence band into a doublet ($\Gamma_{5}$, in-plane
\emph{p}-states) and a singlet ($\Gamma_{1}$, out-of-plane
\emph{p}-states). As a result of the spin-orbit coupling, the
orbital doublet is split into two (orbit+spin) doublets
($\Gamma_{7(5)}$, $\Gamma_{9(5)}$), and the orbital singlet forms a
$\Gamma_{7(1)}$ doublet with a very small admixture of the
$\Gamma_{7(5)}$ states.

Exciton $C$, which is active in $\pi$ polarization, is unambiguously
ascribed to the $\Gamma_{7(1)}$ doublet. The attribution of excitons
$A$ and $B$ is not straightforward. The splitting energies are given
by:\cite{Lamb02}

\begin{equation}
E_{\Gamma_{9(5)}}-E_{\Gamma_{7(5)}}=\frac{\Delta_1
+3\Delta_2}{2}-\sqrt{\left(\frac{\Delta_1
-\Delta_2}{2}\right)^{\lefteqn 2}+2\Delta_3^2}
 \label{Enery9-75}
\end{equation}
\begin{equation}
E_{\Gamma_{9(5)}}-E_{\Gamma_{7(1)}}=\frac{\Delta_1
+3\Delta_2}{2}+\sqrt{\left(\frac{\Delta_1
-\Delta_2}{2}\right)^{\lefteqn 2}+2\Delta_3^2}
 \label{Enery9-71}
\end{equation}

Lambrecht and co-workers\cite{Lamb02} have obtained the following
parameters for pure ZnO: ${\Delta_1=38}$~meV,
${|\Delta_2|=4.53}$~meV and ${\Delta_3=-3.05}$~meV. We cite only the
absolute value of $\Delta_2$, because we do not want to suggest any
particular energy order for the valence states: a positive sign of
$\Delta_2$ implies that the valence state $A$ has the
$\Gamma_{9(5)}$ symmetry,\cite{Langer70,Reynolds99,Gil01a,Gil01b} a
negative sign that $A$ is
$\Gamma_{7(5)}$.\cite{Lamb02,Grub04,Thom60} Actually, taking into
account that the spin-orbit $\Delta_3$ parameter is much smaller
than $(\Delta_1 -\Delta_2)$, we obtain the approximate splitting
energies: $E_{\Gamma_{9(5)}}-E_{\Gamma_{7(5)}}=2\Delta_2$,
$E_{\Gamma_{9(5)}}-E_{\Gamma_{7(1)}}=\Delta_1 + \Delta_2$, and
$E_{\Gamma_{7(5)}}-E_{\Gamma_{7(1)}}=\Delta_1 - \Delta_2$.

In thin epilayers with the c-axis  perpendicular to the plane,
reflectivity measurements in $\pi$ polarization (${E\parallel c}$,
${k\perp c}$) are particularly difficult. In order to observe
exciton $C$ we performed reflectivity measurements with the sample
tilted by an angle of 45 degrees with respect to the optical axis.
Then the selection rule is relaxed, and we can observe all three
excitons [Fig.~\ref{fig:Reflectivity45}(a)]. Spectra are analyzed
using the polariton formalism described in the Appendix.

We observe a systematic increase of the excitonic transition
energies with the Co concentration in all studied samples
[Fig.~\ref{fig:Reflectivity45}(b)]. This indicates an increase of
the energy gap of Zn$_{1-x}$Co$_{x}$O with $x$. The energy of
exciton $A$ is $E_A(x)=3372.6$~meV$~+~x\times1690$~meV. The energy
differences between the excitonic lines in zero field are
$E_B-E_A=7$~meV and $E_C-E_B=47$~meV, whatever the Co content.

Using the approximate splitting energies above, we obtain ${\Delta_1
= 50}$~meV and ${|\Delta_2| = 3.5}$~meV. The difference in
$\Delta_1$ between the present layers and values previously reported
for bulk samples\cite{Lamb02} may be due epitaxial strain, but also
for a good part to the fact that polaritons were not taken into
account.

Finally, as we observe excitons, we should take into account the
electron-hole exchange, which is by no way negligible in ZnO. As the
value of the parameters describing the excitons in ZnO is still a
matter of debate, we keep for a while this simple description, and
we will come back to this important point in
section~\ref{sec:Analysis}.B.

\begin{figure}
\includegraphics*[width=85mm]{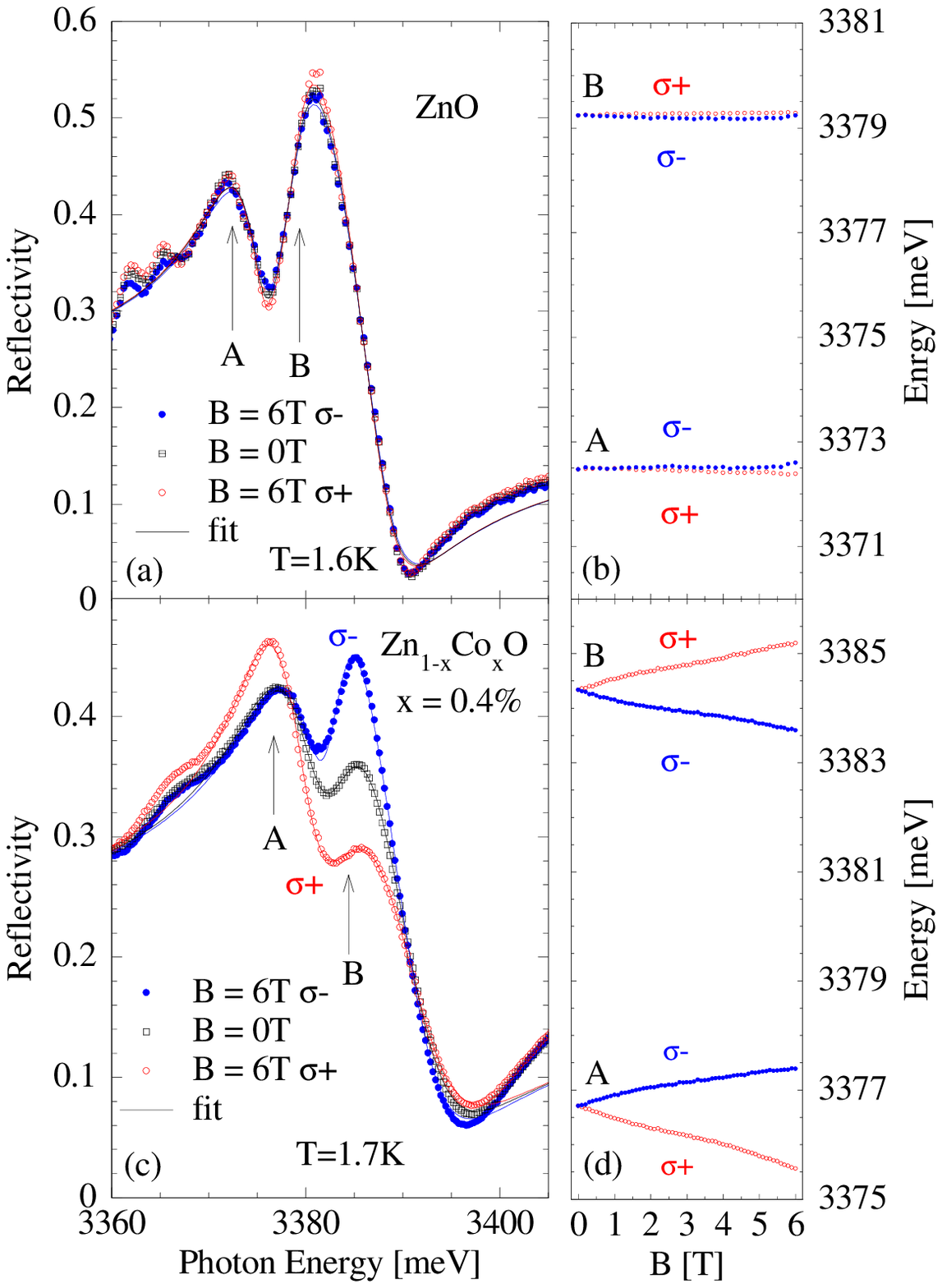}
\caption[]{(color online) (a) Reflectivity spectra of ZnO at
${B=0}$~T and at ${B=6}$~T in $\sigma^+$ and $\sigma^-$ circular
polarizations (symbols). The positions of excitons $A$ and $B$, as
determined from the fit of the ${B=0}$~T spectrum (solid line), are
marked by arrows. The Zeeman effect induces a small, opposite shift
of the two excitons, which changes their overlap and induces a
change of the reflected intensity. (b) The position of excitons $A$
and $B$ in ZnO, as determined from a fits of the reflectivity
spectra shown in (a). (c) Reflectivity spectra of
Zn$_{0.996}$Co$_{0.004}$O at ${B=0}$~T, and ${B=6}$~T in $\sigma^+$
and in $\sigma^-$ polarizations. Solid lines are the fits to the
spectra. (d) The position of excitons $A$ and $B$, versus magnetic
field. At ${B=6}$~T, the value of the splitting is -1.8~meV for
exciton $A$ and 1.6~meV for $B$.} \label{fig:RefSpcZeeman}
\end{figure}

\subsection {Zeeman effect}
We performed magneto-optical spectroscopy in the Faraday
configuration, by applying a magnetic field up to 6~T parallel to
both the c-axis and light propagation. In this configuration,
excitons $A$ and $B$ are observed in both $\sigma^+$ and $\sigma^-$
circular polarizations. The reflectivity spectra in each circular
polarization were analyzed by taking into account the formation of
excitonic polaritons. Polaritons are particulary important in the
case of ZnO because of the strong coupling between excitons and
photons. The dielectric function and the reflectivity spectra near
excitonic resonances have been described in detail by Hopfield and
Thomas\cite{Hopf63} and by Lagois.\cite{Lago77,Lago81} The
application of this theory to our case is described in Appendix.

The transition energies of both $A$ and $B$ and the corresponding
values of the Zeeman splitting [Fig.~\ref{fig:RefSpcZeeman}] where
obtained in a two-step procedure. First, the whole set of parameters
(polarizability $\alpha_{0A,B}$, position $\omega_{A,B}$, width
$\Gamma_{A,B}$, the dead layer thickness $d$, and the non resonant
absorption contribution $i\epsilon'$) were determined by fitting the
zero field spectra. Then, the spectra under magnetic field were
fitted by adjusting only five parameters: $\omega_{A,B}$,
$\Gamma_{A,B}$, and the relative polarizability of excitons $A$ and
$B$ (keeping constant the sum ${\alpha_{0A}+\alpha_{0B}}$).

We did not observe any excitonic diamagnetic shift, as expected due
to the small excitonic Bohr radius (${R_X=18\AA}$) in
ZnO\cite{Blat82} (the diamagnetic shift is smaller than 0.1~meV at
$B=6$~T, so it is smaller than 0.003~meV$/T^2$).

For ZnO, the Zeeman splitting of excitons $A$ and $B$ at 6T remains
very small (about 0.1 meV). Such a value corresponds to an effective
excitonic Land\'e factor less than 0.3. Such a small value of the
Land\'e factors for the allowed excitonic transitions in ZnO has
been already reported in the Voigt configuration by Blattner et
al.,\cite{Blat82} and more recently by Reynolds et al.
\cite{Reynolds99} In this work, the experiments are performed in the
Faraday configuration and they confirm unambiguously the previous
observations.

\begin{figure} 
\includegraphics*[width=85mm]{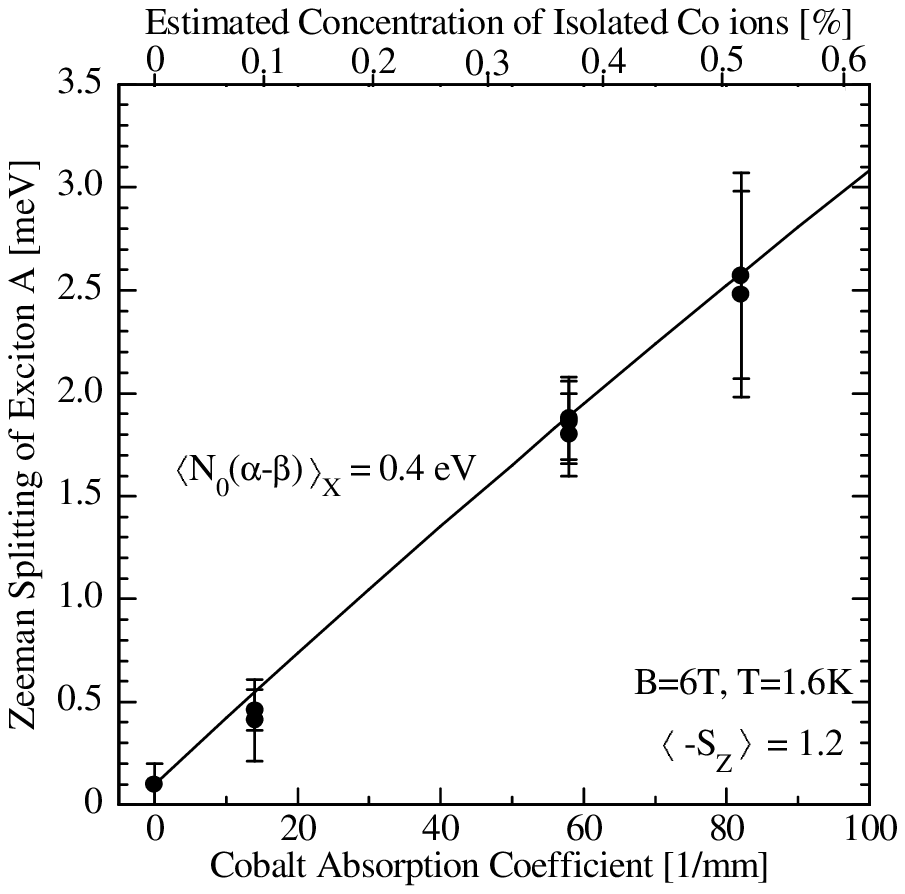}
\caption[]{Zeeman splitting of exciton $A$, measured at ${B=6}$~T
and ${T=1.6}$~K, so that the mean spin of isolated Co is ${\langle
S_z\rangle=1.2}$ [Eq.~(\ref{eq:meanSz})]. The horizontal scale
(bottom scale) is the absorption coefficient of the $^4A_2
\rightarrow\,^2E 2\overline A$ transition of Co (see
Fig.~\ref{fig:Cocalibration}). The top scale (nonlinear) displays
the density of free spins $x_{eff}$ expected for a random
distribution of Co on the Zn sublattice. The solid line shows the
giant Zeeman splitting ${\langle N_0 (\alpha-\beta) \rangle_X}\times
x_{eff}\times \langle \texttt-S_z\rangle$ calculated with an
effective ${\langle N_0 (\alpha-\beta) \rangle_X}=0.4$~eV.}
\label{fig:integrals}
\end{figure}

\begin{figure} 
\includegraphics*[width=75mm]{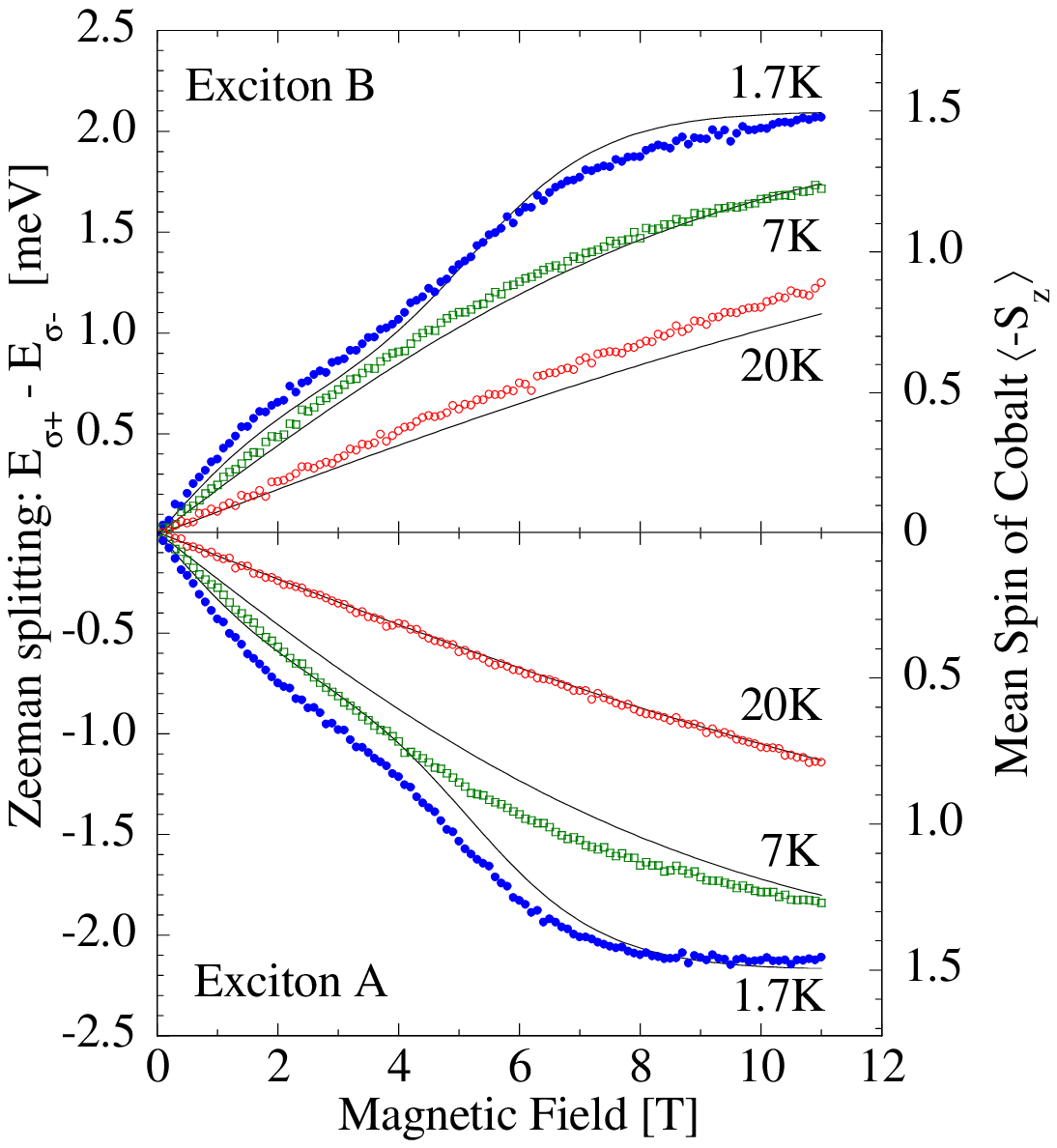}
\caption[]{(color online) Dots and left axis: Zeeman splitting  of
excitons $A$ and $B$ measured at three temperatures; lines, right
axis: mean spin of isolated Co ions as given by
Eq.~(\ref{eq:meanSz}). The sample has ${x_{eff}=0.36\%}$. The left
and right scales are chosen so that ${\langle N_0 (\alpha-\beta)
\rangle_X}=0.40$~eV for exciton $A$ and 0.39~eV for $B$.}
\label{fig:Zeeman}
\end{figure}

In samples doped with cobalt ions, we observe an enhancement of the
excitonic Zeeman splitting. The Zeeman splitting of excitons $A$ and
$B$ is almost opposite. It increases with the Co concentration
[Fig.~\ref{fig:integrals}], and decreases with temperature.
Fig.~\ref{fig:Zeeman} shows the Zeeman splitting of excitons $A$ and
$B$ as a function of the magnetic field up to 11~T, at three
temperatures: 1.7~K, 7~K and 20~K. A saturation appears at the
lowest temperature and the highest magnetic field. The Zeeman
splitting is proportional to the mean spin projection calculated for
isolated Co in section~\ref{sec:Co} [Eqs.~(\ref{eq:meanSz-exp}) and
(\ref{eq:meanSz})].

This enhancement of the Zeeman splitting which increases with the
cobalt content, and its proportionality to the Co magnetization, are
essential for this study. They are analyzed in
section~\ref{sec:Analysis} as a consequence of ${s,p\texttt{-}d}$
interactions.

\begin{figure}
\includegraphics*[width=85mm]{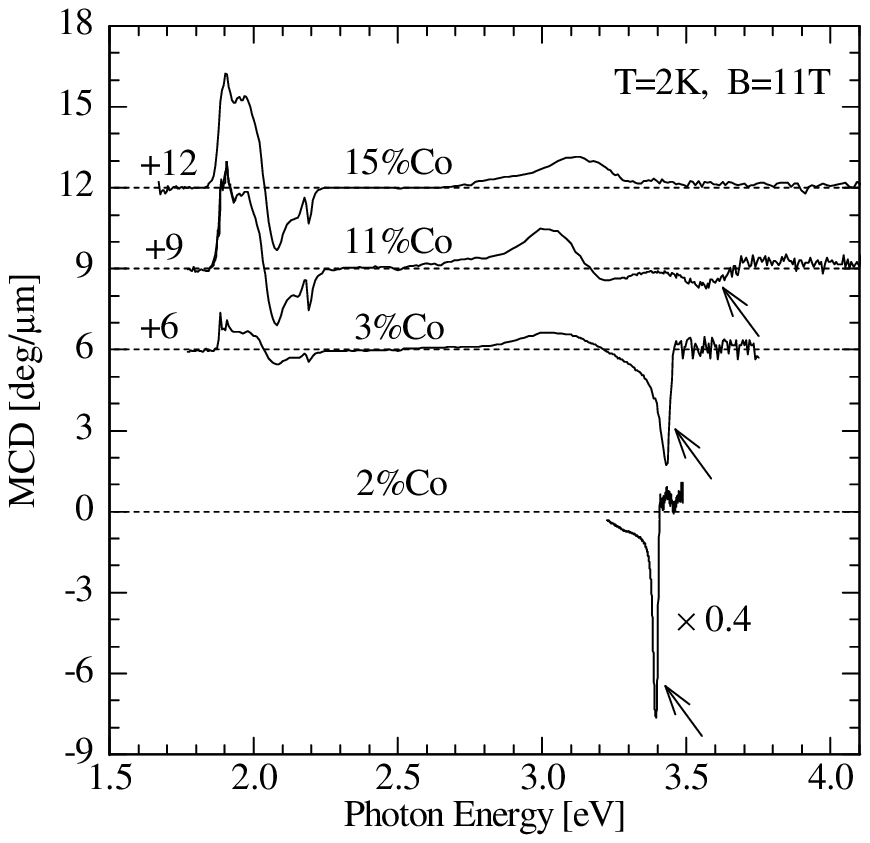}
\caption{(a) Magnetic Circular Dichroism (MCD) obtained in
transmission on four samples with various Co concentrations. The MCD
signal contains three contributions, one (pointed by arrow, most
intense in the sample with the lowest Co concentration, 2\%~Co)
close to the bandgap at 3.4~eV, the second below the
gap,\cite{Schw03} the last one (most intense in the sample with the
highest Co concentration, 15\%~Co) in coincidence with the internal
transitions of Co (1.8~eV - 2.3~eV). Spectra are shifted for
clarity.} \label{fig:mcdspc}
\end{figure} 

\subsection {Magnetic Circular Dichroism}
In samples with Co concentration above $2\%$, the broadening of the
excitonic lines prevents a direct observation of the Zeeman
splitting of excitons $A$ and $B$. In such a case, we measured the
degree of circular polarization of the transmitted light,
${(I_{\sigma+} - I_{\sigma-})/(I_{\sigma+} + I_{\sigma-})}$. For the
sake of comparison with published
data,\cite{Ando01,Ando01b,Ando02,Ando04,Tuan04,Schw03,Kitt05b} we
plot the MCD defined as:

\begin{equation}
\mathrm {MCD}=\frac{180^{\circ}}{4\pi}(k_{\sigma-} - k_{\sigma+}),
 \label{eq:mcd}
\end{equation}
where $k_{\sigma-}$ and $k_{\sigma+}$ are the optical absorption
coefficients in $\sigma^-$ and $\sigma^+$ polarizations. Assuming a
weak absorption and neglecting multiple reflections, the MCD can be
expressed as
\begin{equation}
\mathrm {MCD}=\frac{180^{\circ}}{2\pi l}\frac{I_{\sigma+} -
I_{\sigma-}}{I_{\sigma+} + I_{\sigma-}},
 \label{eq:mcd1}
\end{equation}
where $l$ is the thickness of the sample, and $I_{\sigma+}$ and
$I_{\sigma-}$ the intensities of transmitted light in $\sigma^-$ and
$\sigma^+$ polarizations. The MCD depends on the photon energy and
it is usually very strong near a resonant line or band split by
Zeeman effect. In simple cases, the Zeeman splitting $\Delta E$ can
be estimated \cite{Ando01} from the expression
MCD~$={180^{\circ}/4\pi\times dk/dE\times \Delta E}$, where
${dk/dE}$ is the derivative of the absorption coefficient in zero
field. However, to use properly this equation, the line shape should
remain constant: this is not the case in Zn$_{1-x}$Co$_{x}$O, where
the line shape changes significantly, and where the $A$ and $B$
excitonic transitions are close together. Additionally, they split
in opposite directions, as clearly observed on the reflectivity
spectra of Fig.~\ref{fig:RefSpcZeeman}. These changes of the
lineshape are the reason, why we did not attempt to determine the
absolute value of the Zeeman splitting from these MCD measurements.
In each sample, we only analyzed the temperature, spectral and
magnetic field dependence of the MCD signal.

We observe three main MCD features over a large spectral range, from
1.7~eV to 4~eV (Fig.~\ref{fig:mcdspc}).

The first one is related to the energy gap (3.4~eV). In samples up
to a Co content of 11\%, it is seen as a sharp negative component,
which broadens as the Co content increases. It is accompanied by a
broader positive component from 2.9 to 3.2~meV, which was
attributed\cite{Tuan04,Kitt05b,Schw03} to a charge transfer
transition, or level to band transition. It is this peak which was
used to study the onset of ferromagnetism upon
annealing.\cite{Kitt05b,Schw03}

The third one corresponds to the internal transitions of cobalt, and
it is observed between 1.8~eV and 2.3~eV in samples with a Co
concentration over 0.4\%. Its intensity systematically increases
with the Co concentration. This MCD signal is quite complex and
mainly results from the different intra-ionic absorption lines
analyzed in section~\ref{sec:Co}.

\begin{figure}
\includegraphics*[width=85mm]{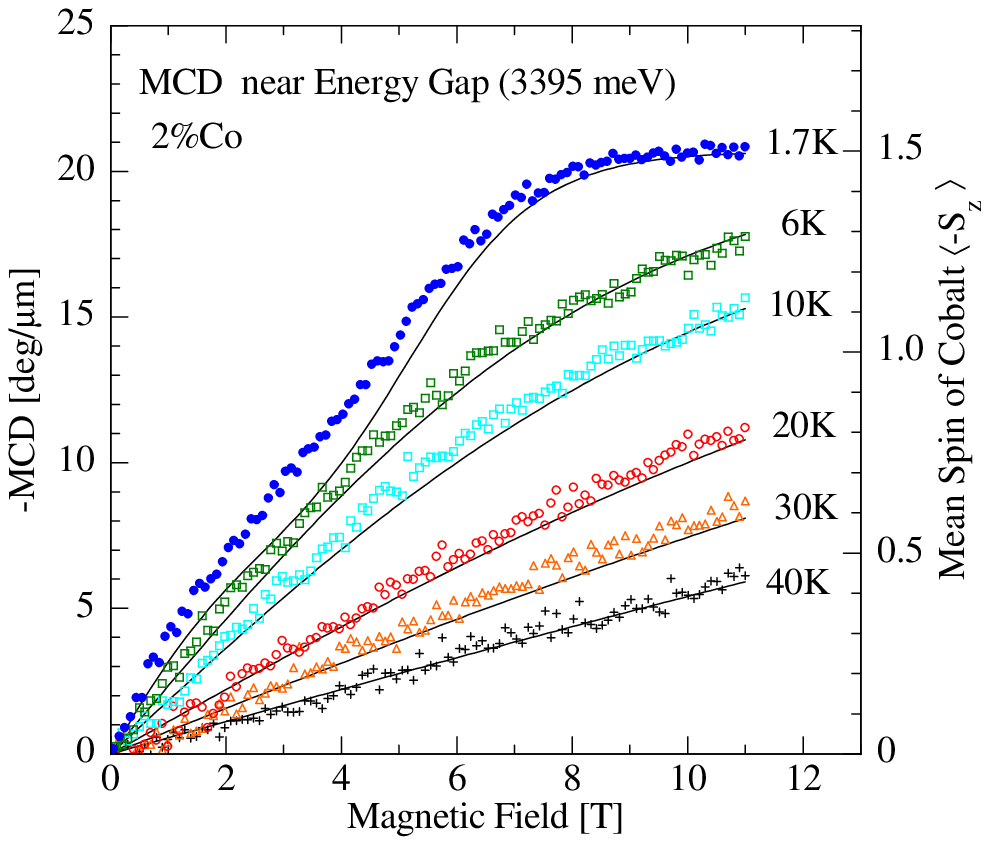}
\caption{ (color online) Symbols, left axis: Magnetic Circular
Dichroism in transmission, at a photon energy 3395~meV, close to the
bandgap; lines, right vertical axis: mean spin of isolated Co ions
[Eq.~(\ref{eq:meanSz})]. The sample is Zn$_{1-x}$Co$_{x}$O with
2\%~Co.} \label{fig:mcd}
\end{figure}

The temperature and magnetic field dependence of the MCD signal near
the energy gap was found to be proportional to that of the magnetic
moment of isolated Co ions [Eq.~(\ref{eq:meanSz})], even in a sample
with a cobalt content of 4.5\% [Fig.~\ref{fig:mcd}]. For higher
concentrations, the resonant signal becomes too weak to be detected
with reasonable accuracy by our experimental setup.

\section {Analysis}
\label{sec:Analysis}

The magneto-optical effects observed near the energy gap and
described in sections \ref{sec:X}.B and \ref{sec:X}.C are now
interpreted as a consequence of the ${s,p\texttt{-}d}$ interaction
between Co ions and band carriers. We first quantify the ion-exciton
coupling (subsection A); then (subsection B) we discuss the effect
of the electron-hole exchange in the exciton; finally (subsection
C), we discuss briefly the magnetic properties of the present
Zn$_{1-x}$Co$_{x}$O layers, as they appear from magneto-optical
spectroscopy.

\subsection {Excitonic Giant Zeeman splitting \\ without electron-hole exchange}
In II-VI DMSs, the presence of the ${s,p\texttt{-}d}$ exchange
interactions leads to large magneto-optical effects (giant Zeeman
splitting and giant Faraday rotation). These effects have been
studied in detail in several wurtzite II-VI DMSs like
Cd$_{1-x}$Mn$_{x}$Se,\cite{Agga83,Arci86}
Cd$_{1-x}$Co$_{x}$Se,\cite{Nawr91,Hamd92}
Cd$_{1-x}$Fe$_{x}$Se,\cite{Twar90} Zn$_{1-x}$Mn$_{x}$Se,\cite{Yu95}
and Cd$_{1-x}$Cr$_{x}$S.\cite{Herb98}

The effective Hamiltonian describing the giant Zeeman effect is
\begin{equation}
H_{ex}^{CB} = -N_0 \alpha x_{eff} \langle \bf{S}\rangle \cdot
\bf{\hat s},
 \label{eq:exchangeH CB}
\end{equation}
for the conduction band, and
\begin{equation}
H_{ex}^{VB} = -N_0 \beta x_{eff} \langle \bf{S}\rangle \cdot
\bf{\hat s},
 \label{eq:exchangeH VB}
\end{equation}
for the valence band. In these expressions, $N_0$ is the number of
cations per unit volume, $x_{eff}$ the free magnetic ion content,
$\alpha$ the exchange constant for the conduction band, $\beta$ the
exchange constant for the valence band, ${\langle
\textbf{S}\rangle}$ the mean spin of the free magnetic ions, in our
case Co ions, and $\bf{\hat s}$ is the spin of the carrier. The mean
value of the isolated Co spin $\langle S_z\rangle$ has been already
obtained as a function of temperature in section III
[Eqs.~(\ref{eq:meanSz-exp}), (\ref{eq:meanSz})], for a magnetic
field applied parallel to the c-axis.

In order to calculate the effect of these effective Hamiltonians, we
must take into account the structure of the valence band. We note
${G_e = 1/2N_0\alpha x_{eff} \langle \texttt-S_z\rangle}$ and ${G_h
= 1/2N_0\beta x_{eff}\langle \texttt-S_z\rangle}$. Neglecting the
contribution of the weak Zeeman effect of the host semiconductor,
the energies of the different excitonic transitions under magnetic
field are given
by:\cite{Agga83,Arci86,Nawr91,Hamd92,Twar90,Yu95,Herb98}
\begin{equation}
E_{\Gamma_{9(5)}}^{\sigma\pm} = E_0 - \Delta_1 -\Delta_2 \mp G_e
\pm G_h
 \label{eq:EneryG9},
\end{equation}
\begin{subequations}
\label{eq:EneryG7}
\begin{equation}
E_{\Gamma_{7(5)}}^{\sigma\pm} = E_0 - \frac{\Delta_1 -\Delta_2}{2}
\pm G_e -E_\pm,
\end{equation}
\begin{equation}
E_\pm = \sqrt{\left(\frac{\Delta_1 -\Delta_2}{2}\pm
G_h\right)^2+2\Delta_3^2},
\end{equation}
\end{subequations}
where $(E_0 -\Delta_1 -\Delta_2)$ is the zero field energy of the
$\Gamma_{9(5)}$ excitonic transition. We do not observe any feature
related to the C exciton ($\Gamma_{7(1)}$) in the configuration
$k\parallel c$, as a consequence of the small spin-orbit coupling in
ZnO [${\Delta_3<<(\Delta_1 -\Delta_2)}$, see section~\ref{sec:X}.A].
Hence we can simplify Eq.~(\ref{eq:EneryG7}) to
$E_{\Gamma_{7(5)}}^{\sigma\pm} = E_0 - \Delta_1 +\Delta_2 \pm G_e
\mp G_h$. Such a simplification leads to opposite Zeeman splittings
for the two optically active excitons :
$E_{\Gamma_{7(5)}}^{\sigma+}-E_{\Gamma_{7(5)}}^{\sigma-}=
-(E_{\Gamma_{9(5)}}^{\sigma+}-E_{\Gamma_{9(5)}}^{\sigma-})= 2(G_e -
G_h) =N_0(\alpha-\beta) x_{eff}\langle \texttt-S_z\rangle$. This is
what we observe experimentally. One consequence however is that the
$\alpha$ and $\beta$ exchange integrals cannot be determined from
the experiment: only their difference can be obtained. Another
consequence is that the contributions of the two active excitons to
the MCD tend to cancel each other in samples where the linewidth is
not small with respect to the ${A\texttt-B}$ splitting.

There is also a direct relationship between the sign of the giant
Zeeman effect and the valence band ordering. From the sign of the
giant Zeeman splitting we observe on $A$ and $B$, a conventional
valence band order ($\Gamma_{9(5)}$, $\Gamma_{7(5)}$,
$\Gamma_{7(1)}$) implies a positive sign of $\alpha-\beta$ (as
usually observed in II-VI DMSs). This ambiguity would disappear in
more concentrated samples, where the excitonic Zeeman splitting will
become comparable to the difference ${(\Delta_1 -\Delta_2)}$, if the
lines remained sharp enough.

In our diluted samples $({x<0.5\%})$, the magnitude of the exchange
integral difference determined on the optically active excitons,
${\langle N_0 (\alpha-\beta) \rangle_X}$, is about 0.4~eV
[Fig.~\ref{fig:integrals}].

In samples where the two excitons are not resolved (such as in our
samples with a few \%~Co), one can attempt to analyze the MCD
phenomenologically, by deducing a Zeeman splitting from a comparison
between the MCD spectra and the derivative of the zero-field
absorption spectrum.\cite{Ando01} We obtain this way (not shown) a
value of the coupling which is of similar magnitude, or slightly
smaller. As noticed above, using the MCD from samples where the two
active excitons are not properly resolved certainly leads to an
underestimation of the coupling, since their contributions tend to
cancel each other. In addition, we will see now that strong effects
are expected from electron-hole exchange within the exciton, when
the giant Zeeman splitting is of the same order as the
${A\texttt-B}$ splitting.

\subsection {${s,p\texttt{-}d}$ interaction in ZnO based DMS\\with electron-hole exchange}

We now include the effect of the electron-hole exchange. We will see
that the effective exchange integrals difference ${\langle N_0
(\alpha-\beta) \rangle_X}$, seen by the excitons can be much smaller
than the difference of the exchange integrals valid for free
carriers, ${N_0(\alpha-\beta)}$. We follow the description given by
B.~Gil et al. when analyzing the effect of the biaxial stress in
epitaxial layers,\cite{Gil01a,Gil01b} and we add the giant Zeeman
effect $G_e$ and $G_h$. The position of the three excitons $A, B,
C$, which can be optically active from symmetry considerations, is
obtained by diagonalizing a $3\times3$ matrix. This matrix mixes the
two excitons active in $\sigma^{+}$ polarization (one formed from
the electron-hole states
${|s\downarrow\rangle~|p^{+}\uparrow\rangle}$, which gives rise to
exciton $A$ if $\Delta_{2}>0$, and
${|s\uparrow\rangle~|p^{+}\downarrow\rangle}$), and the $C$ exciton
state ${|s\downarrow\rangle~|p^{z}\uparrow\rangle}$. It reads (apart
from an overall shift):\cite{Gil01a}

\begin{equation}
\left(%
\begin{array}{ccc}
  -\Delta_2 & \gamma & 0 \\
  \gamma & \Delta_2 & -\sqrt{2}\Delta_3 \\
  0 & -\sqrt{2}\Delta_3 & \tilde{\Delta}_1 -\gamma\\
\end{array}%
\right) \label{eq:Exch}
\end{equation}
where $\tilde{\Delta}_1$ is a trigonal crystal field taking into
account the effect of the trigonal strain,\cite{Gil01a} $\gamma$ is
the electron-hole exchange energy. The giant Zeeman effect adds to
that
\begin{equation}
\left(%
\begin{array}{ccc}
  G_h-G_e & 0 & 0 \\
  0 & G_e-G_h & 0 \\
  0 & 0& G_h-G_e\\
\end{array}%
\right) \label{eq:GZ}
\end{equation}

The same matrix, with opposite giant Zeeman terms, applies in
$\sigma^{-}$ polarization.

The most important effect of the electron-hole exchange within the
exciton, for the present study, is that it mixes the $A$ and $B$
states active in a given circular polarization.

If spin-orbit coupling ($\Delta_{2}$) dominates over exchange
($\gamma$), then one observes the $A$ and $B$ excitons formed on one
of the electron-hole states,
${|s\downarrow\rangle~|p^{+}\uparrow\rangle}$, and
${|s\uparrow\rangle~|p^{+}\downarrow\rangle}$ (respectively for $A$
and $B$ if $\Delta_{2}>0$). These are eigenstates of the c-axis spin
projection operator, so that they are fully affected by the giant
Zeeman effect : In that case, one would measure directly the
spin-carrier coupling ${N_0(\alpha-\beta)}$ on the plot of
Fig.~\ref{fig:integrals}. In the opposite case, the two exciton
states are formed on the symmetric and antisymmetric combinations of
the previous states, resulting in a spin triplet (dark exciton) and
a spin singlet (bright exciton), respectively. Both have vanishing
spin components along the c-axis, so to the first order they do not
exhibit the giant Zeeman effect in the Faraday configuration. As
visible in Fig.~\ref{fig:integrals}, this is not the case, but any
intermediate configuration is possible.

\begin{figure}
\includegraphics*[width=85mm]{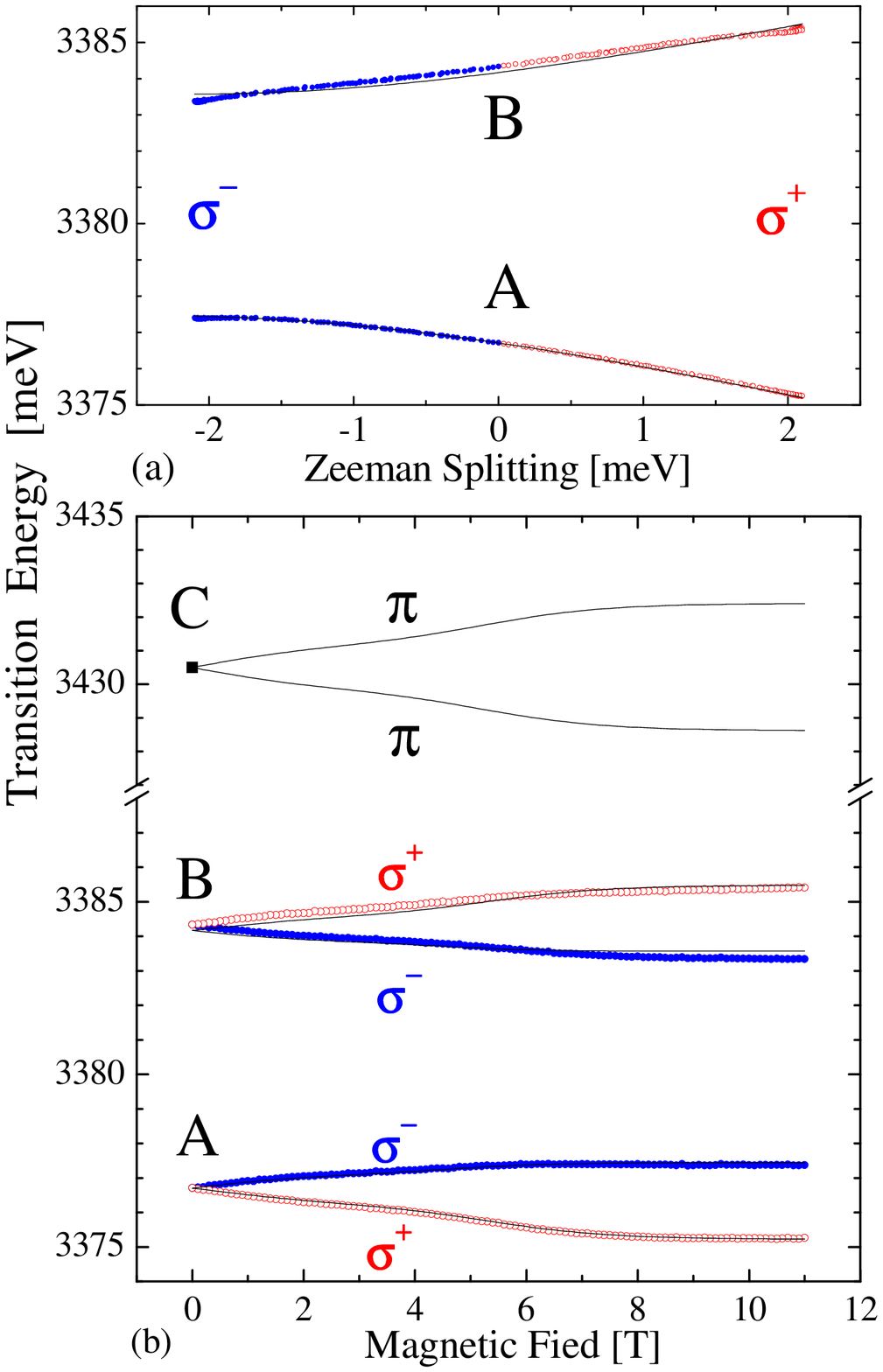}
\caption{(color online) (a) Position of the $A$ and $B$ excitons,
for the same sample as in Fig.~\ref{fig:RefSpcZeeman}b (with
0.4\%~Co), measured at different values of the applied field. The
horizontal axis is the splitting measured on exciton $A$, plotted as
positive if the $\sigma^{+}$ line is at higher energy. (b) Position
of the $A$ and $B$ excitons, as a function of the applied field, and
of exciton $C$ at zero field, for the same sample as in
Fig.~\ref{fig:RefSpcZeeman}b (with 0.4\%~Co). Symbols are
experimental data (blue for $\sigma^{-}$, red for $\sigma^{+}$),
lines are calculated (including for exciton $A$).}
\label{fig:anticross}
\end{figure}

Fig.~\ref{fig:anticross}(a) displays the position of the $A$ and $B$
excitons, from Fig.~\ref{fig:RefSpcZeeman}(b), as a function of the
splitting of exciton $A$. It is clear that we observe an
anticrossing between the two excitons. It is reasonable to ascribe
this anticrossing the the electron-hole interaction, and from the
minimum distance, the value of $\gamma$ is a few meV. However, the
two excitons do not show the same curvature, which suggests that we
cannot restrict ourself to the interaction between $A$ and $B$ only.

Fig.~\ref{fig:anticross}(b) compares the experimental position
(symbols) of excitons $A$ and $B$, as a function of the applied
field, and of exciton $C$, at zero field, to the calculated
eigenvalues of the Hamiltonian [Eqs~(\ref{eq:Exch}) and
(\ref{eq:GZ}), solid lines]. The average spin projection of Co ions
is calculated using Eq~(\ref{eq:meanSz}). A good fit is obtained for
two sets of values of the parameters (which cannot be
distinguished at the scale of the figure):\\
Set~1: ${\tilde{\Delta}_1=51~\textrm{meV}}$,
${\Delta_2=3.0~\textrm{meV}}$, ${\Delta_3=6.3~\textrm{meV}}$,
${\gamma=3.4~\textrm{meV}}$, ${N_0(\alpha-\beta)=0.8~\textrm{eV}}$,
\\or \\
Set~2: ${\tilde{\Delta}_1=53~\textrm{meV}}$,
${\Delta_2=-2.1~\textrm{meV}}$, ${\Delta_3=0.1~\textrm{meV}}$,
${\gamma=3.1~\textrm{meV}}$,
${N_0(\alpha-\beta)=-0.8~\textrm{eV}}$.\\
The value of $\tilde{\Delta}_1$ corresponds to a trigonal stress of
13~kbar, according to Ref.~\onlinecite{Gil01a}.

In both cases, we obtain a very good fit of the lowest exciton, and
the fit for the second optically active exciton is reasonable,
although not as good. The splittings of both excitons are almost
opposite, as experimentally observed (note that if we take the
parameters from Ref.~\onlinecite{Gil01b}, the  splitting calculated
for $B$ almost vanishes). We obtain no information about the valence
band ordering (a similar conclusion was claimed by Lambrecht et
al.\cite{Lamb02} in their analysis of the effect of strain). This is
due to the fact that we mainly observe an anticrossing between the
two excitons $A$ and $B$, which is a symmetrical process. Hence, our
fit is mainly sensitive to the relative value of ${(G_e-G_h)}$ and
$\gamma$. We may simply notice that the hypothesis ${\Delta_2}>0$
leads us to a much more isotropic spin-orbit coupling.

The value obtained for the spin-carrier coupling,
$N_0|\alpha-\beta|\simeq$~0.8~eV, has to be considered as a first
estimate. A larger value of the electron-hole exchange $\gamma$
reduces the sensitivity of the exciton to the spin splitting,
implying a larger spin-carrier coupling. It should be particularly
interesting to measure the excitonic splitting on samples with a
higher Co content.

This value is smaller than in many other DMSs. For example, in
Cd$_{1-x}$Co$_{x}$Se, another wurtzite Co-based DMS, the exchange
integral difference is 2.1~eV.\cite{Nawr91,Hamd92} However,
comparable or even smaller values of $N_0|\alpha-\beta|$ have been
also reported in wurtzite DMSs, for example -0.26 eV in
Cd$_{1-x}$Cr$_{x}$S.\cite{Herb98} It is also in significant
disagreement with theoretical predictions. Blinowski and
co-workers\cite{Blin02} have calculated ${N_0\beta = -3.2}$~eV in
the case of Zn$_{1-x}$Mn$_{x}$O. They have also pointed out that the
absolute value of $N_0\beta$ is expected to be larger for Co than
for Mn ions. Since $N_0\alpha$ is usually about $0.25$~eV and does
not depend strongly on the host material,\cite{Liu61} one could
expect ${N_0(\alpha-\beta)>3.4}$~eV in the case of
Zn$_{1-x}$Co$_{x}$O. A large exchange integral,
${N_0(\alpha-\beta)=2.3}$~eV, has been proposed for
Zn$_{1-x}$Co$_{x}$O nanocristallites,\cite{Schw03} but the details
of the spectra supporting this conclusion were not given. If we
assume a positive sign of $N_0\beta$ (which is not the usual case
but is not excluded by theory, e.g., in
Cd$_{1-x}$Cr$_{x}$S),\cite{Herb98} and ${N_0\alpha=0.25}$~eV, then
we obtain $N_0\beta$ about 1~eV. Such a small value puts severe
restrictions to the possible influence of carriers on
ferromagnetism, at least through Zener-like mechanisms.

\subsection {Magnetism of Zn$_{1-x}$Co$_{x}$O}
A measure of the excitonic Zeeman splitting and of the MCD signal
near the energy gap is known to represent an efficient method to
study the magnetic properties of DMSs. The dependence on the
magnetization results from ${s,p\texttt{-}d}$ interactions, which
are strong with substitutional magnetic ions. These magneto-optical
properties are less affected by the presence of secondary phases or
inclusions, which fully contribute to the magnetization as measured
by SQUID. Hence the observation of a clear effect of the
magnetization on the excitonic or band-to-band magneto-optical
properties of the semiconductor is a good hint, that the observed
magnetization is due to the DMS.

In the whole set of studied samples, both the Zeeman splitting and
the near-bangap MCD signal agree with the idea of a paramagnetic
system of isolated Co impurities (Maxwell-Boltzmann occupancy of the
$|$$\pm$$\frac32\rangle$ and $|$$\pm$$\frac12\rangle$ spin states of
Co$^{2+}$ [see Eq.~(\ref{eq:meanSz})].

This model is based on the observation of characteristic
\emph{d}-\emph{d} transitions of the $3d^{4}$ electronic
configuration of Co, which confirms the values of the parameters
describing the anisotropy and Zeeman effect in the ground state. The
intensity of these lines increases linearly with the Co content,
which suggests that all Co ions are incorporated in substitution of
Zn. A Maxwell-Boltzmann distribution over the spin components in the
ground state well accounts for the intensity of the ${d\texttt-d}$
transitions, and for the giant Zeeman effect of the optically active
excitons. This demonstrates that Zn$_{1-x}$Co$_{x}$O is a DMS with a
coupling between the bands of the semiconductor and the localized
spins.

It also stresses some differences with respect to the most usual
case of Mn-based II-VI DMSs. The magnetization of the single Co ion
is not described by the isotropic Brillouin function used for cubic
II-VI DMS such as Cd$_{1-x}$Mn$_{x}$Te. The anisotropy causes a step
in the field dependence of the magnetization at low temperature, and
a deviation from the Curie law in the thermal dependence of the
low-field susceptibility. On samples with a few \%~Co, we did not
need to introduce an effective temperature ${T_{eff}=T+T_0}$ in
order to describe the influence of spin-spin interactions between
neighboring spins (at least, beyond the nearest-neighbor spins). Any
such phenomenological Curie-Weiss temperature $T_0$ has to be
smaller than 0.5~K for concentration up to 5\%.

On the other hand, the model assumes isolated Co ions, so that the
eventual magnetic moments associated with Co pairs and complexes are
significantly smaller than the moment associated with isolated Co
ions. The negligible contribution of anti-ferromagnetically coupled
nearest neighbor pairs at low temperature can be explained by short
range superexchange interaction integrals higher than a few tens of
Kelvin.

Contrary to the resonant peak close to the energy gap, the MCD
signal related to internal Co transitions is not necessarily
proportional to the Co magnetization. This signal is governed by the
occupancy of the cobalt spin sublevels, and the Zeeman shifts and
selection rules of the multiline absorption structure.
Experimentally, its temperature dependence is also much weaker than
the corresponding one for the energy gap resonant MCD peak.

In any case, we have not found any evidence of a ferromagnetic
behavior.  Upon increasing the Co content, the excitonic Zeeman
splitting or the MCD continue to follow the paramagnetic behavior
expected for isolated Co impurities - with perhaps the exception of
a small additional contribution below 5~T at our lowest temperature,
which tends to increase with the Co content (compare
Figs.~\ref{fig:Zeeman} and \ref{fig:mcd}). No hysteretic behavior
was detected. This contrasts with other observations on samples
which have been grown by a different method, which exhibit
ferromagnetic magnetization with an easy axis along the c-axis (an
anisotropy which is opposite to the one we see for isolated,
substitutional, $3d^{4}$ cobalt),\cite{Rode03} or with samples which
have been annealed.\cite{Schw03} However, it fully agrees with a
study of the present (or similar) samples by magnetometry and
electronic paramagnetic resonance.\cite{Pasc04}

In a study of the magneto-optical properties of such samples, one
should keep in mind the trend exemplified in Figs.~\ref{fig:Zeeman}
and \ref{fig:mcd} (the presence of an additional contribution to the
MCD signal below 5~T and 1.7~K), and also the strong effect of
electron-hole exchange in the exciton.


\section {Conclusions}
We observe absorption bands and lines due to Co$^{2+}$ intra-ionic
${d\texttt-d}$ transitions. The intensity of
$^4A_2\rightarrow\,^2E\overline E$ is proportional to the total Co
concentration. We observe reflectivity structures related to $A$,
$B$ and $C$ excitons. Their energies increase with Co concentration.
We measured the exchange Zeeman splittings of $A$ and $B$ excitons,
which increase with Co concentration and decrease with temperature.
These exchange Zeeman splittings and the intensity of energy gap
resonant MCD signal are described by using a model based on thermal
occupation of $|$$\pm$$\frac32\rangle$ and $|$$\pm$$\frac12\rangle$
spin levels of purely paramagnetic Co$^{2+}$ ions. A first
estimation of spin-exciton exchange integrals ${N_0|\alpha-\beta|}$
was deduced, which results in a higher value of the spin-carrier
integrals, of the order of 0.8~eV, when electron-hole exchange
interaction is taken into account.

\acknowledgements

We wish to acknowledge helpful discussions with A.~Golnik, P.~Sati,
and A.~Stepanov. We would like to thank also M.~La\"{u}gt,
J.-M.~Chauveau and P.~Venn\'egues for the structural
characterization of the samples (x-ray and Transmission Electron
Microscopy). This work was partially supported by Polish Committee
for Scientific Research (grants 2P03B 002 25 and
PBZ-KBN-044/P03/2001), program Polonium, Center of Excellence CEMOS
(G5MA-CT-2002-04062) and Thematic Network SOXESS
(G5RT-CT-2002-05075).


\appendix*
\section{Reflectivity due to excitonic polaritons}
We present here the details of the polariton model used to fit
reflectivity spectra. The model takes into account the strong
coupling between excitons and photons in ZnO (the longitudinal
transverse splitting, of several meV, is of the order of the
excitonic linewidth). The dielectric function and reflectivity
spectra near excitonic resonances are described in detail by
Hopfield and Thomas\cite{Hopf63} and by Lagois.\cite{Lago77,Lago81}
Only $A$ and $B$ excitons were taken into account. The dispersion of
the polariton branches associated with a propagation of the light
parallel to the c-axis was calculated numerically by solving of the
following equations (\ref{eq:DielectricFuncktion}) and
(\ref{eq:epsilon}):

\begin{eqnarray}
\nonumber \epsilon(k,\omega) = \epsilon_\infty +
\frac{4\pi\alpha_{0A}\omega_A}{\omega_A^2-\omega^2+(\hbar k^2
\omega_A /m^*)-i\omega\Gamma_A} \\ +
\frac{4\pi\alpha_{0B}\omega_B}{\omega_B^2-\omega^2+(\hbar k^2
\omega_B /m^*)-i\omega\Gamma_B},
 \label{eq:DielectricFuncktion}
\end{eqnarray}
\begin{equation}
\epsilon(k,\omega) = k^2c^2/\omega^2.
 \label{eq:epsilon}
\end{equation}
where ${\epsilon_\infty=6.2}$ is the background dielectric constant
of ZnO,\cite{Sega67} $\alpha_{0A,B}$ is the polarizability,
$\Gamma_{A,B}$ are dumping parameters, $\omega_{A,B}$ are the
energies of excitonic transitions,  and ${m^*=0.87m_e}$ is the
effective mass of excitons.\cite{Humm73}

We note $\epsilon_1$, $\epsilon_2$, $\epsilon_3$ the three solutions
of (\ref{eq:DielectricFuncktion}) and (\ref{eq:epsilon}). For each
solution, we can determine the associated refractive index
${n_i=\sqrt{\epsilon_i}}$, the wave vector ${k_i=n_i\omega/c}$ and
the polarization of $A$ and $B$ excitons:
\begin{equation}
\epsilon_i^{A,B}=\frac{4\pi\alpha_{0A,B}\omega_{A,B}}{\omega_{A,B}^2-\omega^2+(\hbar
k_i^2 \omega_{A,B} /m^*)-i\omega\Gamma_{A,B}}\\
\label{eq:PolExcitons}
\end{equation}

The reflection coefficient was derived from surface boundary
conditions for electric field (\ref{eq:E_BC}), magnetic field
(\ref{eq:B_BC}) and the total polarization of $A$ (\ref{eq:exA_BC})
and $B$ (\ref{eq:exB_BC}) excitons:
\begin{equation}
E_I + E_R = E_1 + E_2 + E_3,
 \label{eq:E_BC}
\end{equation}
\begin{equation}
E_I - E_R = n_1E_1 + n_2E_2 + n_3E_3,
 \label{eq:B_BC}
\end{equation}
\begin{equation}
\epsilon_1^AE_1 + \epsilon_2^AE_2 + \epsilon_3^AE_3 = 0,
 \label{eq:exA_BC}
\end{equation}
\begin{equation}
\epsilon_1^BE_1 + \epsilon_2^BE_2 + \epsilon_3^BE_3 = 0.
 \label{eq:exB_BC}
\end{equation}
$E_I$ is the electric field amplitude of incoming light, $E_R$ is
the corresponding one for reflected light, $E_1$, $E_2$ and $E_3$
are the electric field amplitudes of the three polaritons
propagating inside the sample. From (\ref{eq:exA_BC}) and
(\ref{eq:exB_BC}), we get:
\begin{equation}
\frac{E_2}{E_1} = \frac{\epsilon_1^B\epsilon_3^A -
\epsilon_1^A\epsilon_3^B}{\epsilon_2^A\epsilon_3^B -
\epsilon_2^B\epsilon_3^A}
\end{equation}
and
\begin{equation}
\frac{E_3}{E_1} = \frac{\epsilon_1^B\epsilon_2^A -
\epsilon_1^A\epsilon_2^B}{\epsilon_3^A\epsilon_2^B -
\epsilon_3^B\epsilon_2^A.}
\end{equation}
Next, we obtain an effective refractive index $n^\dagger$
\begin{equation}
n^\dagger =\frac{E_I - E_R}{E_I + E_R}= \frac{n_1 +
n_2\frac{E_2}{E_1} + n_3\frac{E_3}{E_1}}{1 + \frac{E_2}{E_1} +
\frac{E_3}{E_1}}
 \label{eq:n_eff1}
\end{equation}

With the derived expression for $n^\dagger$, we can calculate the
reflection coefficient of the layer ${R=|\frac{n^\dagger
-1}{n^\dagger +1}|^2}$, which takes into account the presence of the
polariton branches. However, it is commonly assumed that the
formation of excitons is not possible within a few nanometers close
to the surface. A dead (exciton-free) layer with a thickness $d$
(about two times the excitonic Bohr radius), and a background
refractive index ${n=\sqrt{\epsilon_0}}$ are usually added to the
model. The reflectivity of the bilayer structure leads to an
effective refractive index given by :
\begin{equation}
n^* = n\left
[\frac{(n^\dagger+n)e^{ikd}-n+n^\dagger}{(n^\dagger+n)e^{ikd}+n-n^\dagger}\right]
 \label{eq:n_eff2}
\end{equation}
Finally, the reflectivity spectra are given by :
\begin{equation}
R=\left|\frac{n^* -1}{n^* +1}\right|^2
 \label{eq:R}
\end{equation}

In our samples, the reflectivity spectra are slightly modified by
additional non-resonant absorption. We obtained a better fit of the
experimental spectra by adding an imaginary part $i\epsilon'$ to the
background dielectric constant $\epsilon_\infty$ in Eq
(\ref{eq:DielectricFuncktion}). Examples of the fits of the
reflectivity spectra of ZnO and Zn$_{1-x}$Co$_{x}$O epilayers are
shown in Fig.~\ref{fig:RefSpcZeeman}.

\end{document}